\documentclass[conference]{IEEEtran}
\IEEEoverridecommandlockouts

\ifCLASSINFOpdf
   \usepackage[pdftex]{graphicx}
   \DeclareGraphicsExtensions{.pdf,.jpeg,.png}
\else
   \usepackage[dvips]{graphicx}
   \DeclareGraphicsExtensions{.eps}
\fi

\usepackage[cmex10]{amsmath}

\usepackage[named]{algo}
\usepackage{algorithm} 
\usepackage[noend]{algorithmic}

\usepackage{array}

\usepackage{url}

\usepackage[utf8]{inputenc}
\usepackage{amsfonts}
\usepackage{amssymb,amsmath,amsthm}
\newtheorem{definition}{Definition}
\newtheorem{lemma}{Lemma}
\newtheorem{theorem}{Theorem}

\usepackage{todonotes}

\renewcommand{\AA}{\mathcal{A}}

\newcommand{\PSPACE}{\mathrm{PSPACE}}
\newcommand{\EXPTIME}{\ensuremath{\mathrm{EXPTIME}}}
\newcommand{\BB}{\mathcal{B}}
\newcommand{\CC}{\mathcal{C}}
\newcommand{\JJ}{\mathcal{J}}
\newcommand{\DD}{\mathcal{D}}

\newcommand{\RRR}{\mathcal{R}}
\newcommand{\LL}{\mathcal{L}}
\newcommand{\HH}{\mathcal{H}}

\newcommand{\NN}{\mathbb{N}}
\newcommand{\nNN}{_{n\in\NN}}

\newcommand{\Cl}{{\bf Cl}}
\newcommand{\MM}{\mathcal{M}}
\newcommand{\prob}[1]{\mathbb{P}_{#1}}

\newcommand{\set}[1]{\{ #1 \}}
\newcommand{\val}[1]{\text{val}(#1)}

\newcommand{\monoid}{\mathcal{G}}
\newcommand{\monoidext}{\mathcal{G}_+}

\newcommand{\tendsto}[1]{\mathop{\longrightarrow}_{n}}

\newcommand{\limu}{{\bf u}}
\newcommand{\limv}{{\bf v}}
\newcommand{\limw}{{\bf w}}
\newcommand{\lima}{{\bf a}}
\newcommand{\limb}{{\bf b}}
\newcommand{\lime}{{\bf e}}

\newcommand{\pmin}{p_{\min}}
\newcommand{\explicitnumjclasses}{2^{2|Q|^2}}
\newcommand{\numjclasses}{J}
\newcommand{\mmaxdepth}{{3\cdot \numjclasses^2 }}
\newcommand{\dabound}{\pmin^{2 ^\mmaxdepth}}

\newcommand{\merge}{\textrm{merge}}
\newcommand{\chck}{\textrm{check}}
\newcommand{\apply}{\textrm{apply}}
\newcommand{\finish}{\textrm{finish}}
\newcommand{\wait}{\textrm{wait}}

\newtheorem*{problem}{Problem}
\newtheorem{proposition}{Proposition}

\begin{document}

\title{Deciding the Value 1 Problem of Probabilistic Leaktight Automata}

\author{
\IEEEauthorblockN{Nathana\"el Fijalkow}
\IEEEauthorblockA{\'ENS Cachan, LIAFA,\\ Universit{\'e} Denis Diderot Paris 7, France\\
Email: nath@liafa.jussieu.fr}
\and
\IEEEauthorblockN{Hugo Gimbert\thanks{This work was supported by
French CNRS-PEPS Project ``Stochastic Games and Verification''.}}
\IEEEauthorblockA{CNRS, LaBRI\\ Universit{\'e} de Bordeaux, France\\
Email: hugo.gimbert@labri.fr}
\and
\IEEEauthorblockN{Youssouf Oualhadj}
\IEEEauthorblockA{LaBRI\\ Universit{\'e} de Bordeaux, France\\
Email: youssouf.oualhadj@labri.fr}
}

\maketitle

\begin{abstract}
The value $1$ problem is a decision problem for probabilistic automata over finite words:
given a probabilistic automaton $\AA$,
are there words accepted by $\AA$ with
probability arbitrarily close to $1$?

This problem was proved undecidable recently.
We sharpen this result, showing that the undecidability holds
even if the probabilistic automata have only one probabilistic transition.

Our main contribution is to introduce a new class of probabilistic automata,
called \emph{leaktight automata}, for which the value $1$
problem is shown decidable (and $\PSPACE$-complete).
We construct an algorithm based on the computation of a monoid abstracting 
the behaviors of the automaton, and rely on algebraic techniques developed by Simon
for the correctness proof.
The class of leaktight automata is decidable in $\PSPACE$,
subsumes all subclasses of probabilistic automata
whose value $1$ problem is known to be decidable
(in particular deterministic automata),
and is closed under two natural composition operators.
\end{abstract}

\IEEEpeerreviewmaketitle

\maketitle

\section*{Introduction}

\paragraph*{Probabilistic automata}
Rabin invented a very simple yet powerful
model of probabilistic machine called probabilistic automata,
which, quoting Rabin,
``are a generalization of finite deterministic
automata''~\cite{rabinsem}.
A probabilistic automaton has a finite set of states $Q$ and reads input words
over a finite alphabet $A$. 
The computation starts from the initial state $i$ and consists in reading the input word sequentially;
the state is updated according to transition probabilities 
determined by the current state and the input letter. 
The probability to accept a finite input word is the probability to
terminate the computation in one of the final states $F\subseteq Q$.

From a language-theoretic perspective,
several algorithmic properties of probabilistic automata are known:
while language emptiness is undecidable~\cite{bertoni1,GO10,pazbook},
language equivalence is decidable~\cite{CortesMR07,schutz,Tzeng:1992}
as well as other properties~\cite{CondonL89,CortesMRR08}.

Rather than formal language theory,
our initial motivation for this work
comes from control and game theory:
we aim at solving algorithmic questions about partially
observable Markov decision processes and stochastic games.
For this reason, we consider
probabilistic automata as machines controlled by a blind controller,
who is in charge of choosing the sequence of input letters 
in order to maximize the acceptance probability.
While in a fully observable Markov decision
process the controller can observe the current state of the process to choose adequately 
the next input letter, 
a blind controller does not observe anything
and its choice depends only on the number of letters already chosen.
In other words, the strategy of a blind controller is an input word of the
automaton.

\paragraph*{The value of a probabilistic automaton}
With this game-theoretic interpretation in mind,
we define the \emph{value} of a probabilistic
automaton as the supremum of acceptance probabilities over all input words,
and we would like to compute this value.
Unfortunately, as a consequence of an undecidability result due to Paz,
the value of an automaton is not computable in general.
However, the following decision
problem was conjectured by Bertoni to be decidable~\cite{bertoni1}:

\smallskip

{\bf Value $1$ problem:}
\emph{Given a probabilistic automaton, does the automaton have value $1$?
In other words are there input words whose acceptance probability
is arbitrarily close to $1$?
}

\smallskip

Actually, Bertoni formulated the value $1$ problem in a different yet
equivalent way: ``Is the cut-point $1$ isolated or not?''.
There is actually a close relation between the value $1$ problem
and the notion of isolated cut-point introduced by Rabin
in the very first paper about probabilistic automata.
A real number $0 \leq \lambda \leq  1$ is an \emph{isolated cut-point} if there
exists a bound $\epsilon >0$ such that the acceptance probability of any word is
either greater than $\lambda + \epsilon$ or smaller than $\lambda - \epsilon$.
A theorem of Rabin states that if the cut-point $\lambda$ is isolated,
then the language $L_{\lambda} = \set{w \mid \prob{\AA}(w) \geq \lambda}$ is regular~\cite{rabinsem}.
The value $1$ problem can be reformulated in term of isolated cut-point:
an automaton has value $1$ if and only if $1$ is not an isolated cut-point.
Bertoni proved that for $\lambda$ strictly between $0$ and $1$,
the isolation of $\lambda$ is undecidable in general,
and left the special case $\lambda \in \{0,1\}$ open.

\smallskip

Recently, the second and third authors of the present paper proved that
the value $1$ problem is undecidable~\cite{GO10} as well.
However, probabilistic automata, and more generally partially observable Markov
decision processes and stochastic games,
are a widely used model of probabilistic machines used in many fields 
like software verification~\cite{BBG08,chdr07}, image processing~\cite{images},
computational biology~\cite{biology} and speech processing~\cite{linguistics}.
As a consequence, it is crucial to understand which decision problems are algorithmically tractable
for probabilistic automata.

\paragraph*{Our result}
As a first step, we sharpen the undecidability result:
we prove that the value $1$ problem is undecidable even 
for probabilistic automata with only one probabilistic transition.
This result motivated the introduction of a new class of probabilistic
automata, called \emph{leaktight automata}, for which the value $1$ problem is
decidable.
This subclass subsumes all known subclasses of probabilistic automata sharing this decidability property
and is closed under parallel composition and synchronized product.
Our algorithm to decide the value $1$ problem computes in polynomial space a finite
monoid whose elements are directed graphs and checks whether
it contains a certain type of elements that are value $1$ witnesses.

\paragraph*{Related works}
The value $1$ problem was proved decidable
%by the second and third author of this paper~\cite{GO10}
for a subclass of probabilistic automata called $\sharp$-acyclic
automata~\cite{GO10}.
%The first step towards deciding the value $1$ problem was achieved 
%by $\sharp$-acyclic automata~\cite{GO10}. 
%Since this subclass are a
Since the class of $\sharp$-acyclic automata
is strictly contained in the class of
%is subclass is a
%strict subclass of 
leaktight automata, the result of the present paper
extends the decidability result of~\cite{GO10}.
Chadha et al.~\cite{ChadhaSV09} recently introduced the class
of hierarchical probabilistic automata, 
which is also strictly contained in the class of leaktight automata.
As a consequence of our result, the value $1$ problem is decidable
for hierarchical probabilistic automata.
Our proof techniques totally
depart from the ones used in~\cite{ChadhaSV09,GO10}.
Instead, we make use of algebraic techniques and in particular
Simon's factorization forest theorem, which was successfully used to prove
the decidability of the boundedness problem for distance
automata~\cite{simontropical}.

\paragraph*{Outline}
We give the basic definitions in Section~\ref{sec:def}.
As a first step we present our algorithm to decide the value $1$ problem of probabilistic leaktight automata 
in Section~\ref{sec:dec},
which is followed by the decidability of the leaktight property in Section~\ref{sec:leak}.
Next, in Section~\ref{sec:correct}, we present and prove the technical core of the paper, called the lower bound lemma.
Finally, Section~\ref{sec:ex} investigates properties and provides examples of leaktight automata.
The proofs can be found in the appendix.

\section{\label{sec:def}Definitions}

\subsection{Probabilistic automata}

Let $Q$ be a finite set of states. A probability distribution over $Q$ is a row vector
$\delta$ of size $|Q|$ whose coefficients are real numbers from the interval
$[0,1]$ and such that $\sum_{q \in Q} \delta(q) = 1$.
A probabilistic transition matrix $M$ is a square matrix in $[0,1]^{Q\times Q}$
such that every row of $M$ is a probability distribution over $Q$.

\begin{definition}[Probabilistic automata]
A probabilistic automaton $\AA$ is a tuple $(Q, A, (M_a)_{a \in A}, i, F)$,
where $Q$ is a finite set of states, $A$ is the finite input alphabet, 
$(M_a)_{a \in A}$ are the probabilistic transition matrices,
$i\in Q$ is the initial state and $F\subseteq Q$ is the set of accepting states.
\end{definition}

For each letter $a \in A$, $M_a(s,t)$ is the probability to go from state $s$ to state $t$ 
when reading letter $a$.
Given an input word $w\in A^*$,
we denote by $w(s,t)$ the probability to go from state
$s$ to state $t$ when reading the word $w$.
Formally, if $w = a_1 a_2 \cdots a_n$ then $w(s,t) = (M_{a_1} \cdot M_{a_2} \cdots M_{a_n})(s,t)$.
Note that $0 \leq w(s,t) \leq 1$, for all words $w$ and states $s$ and $t$.
Furthermore, the definition of a probabilistic transition matrix implies that:
%Hugo: our automata are complete: is 'complete' used as a standard? We use
% complete for monoid already
$\sum_{t \in Q} w(s,t) = 1$ for all states $s$.

\begin{definition}[Value and acceptance probability]
The \emph{acceptance probability} of a word $w \in A^*$ by $\AA$ is
$\prob{\AA}(w) = \sum_{f\in F} w(i,f)$.
The \emph{value} of $\AA$, denoted $\val{\AA}$,
is the supremum of the acceptance probabilities over all possible input words:
\begin{equation}
\label{eq:value}
\val{\AA} = \sup_{w \in A^*} \prob{\AA}(w)\enspace.
\end{equation}
\end{definition}

\subsection{The value $1$ problem for probabilistic automata}

We are interested in the following decision problem:

\begin{problem}[Value $1$ Problem]\label{prob:value1}
Given a probabilistic automaton $\AA$, decide whether $\val{\AA} = 1$.
\end{problem}

The value $1$ problem can be reformulated using the notion
of \emph{isolated cut-point} introduced by Rabin in his seminal
paper~\cite{rabinsem}: an automaton has value $1$ if
and only if the cut-point $1$ is \emph{not} isolated.

Whereas the formulation of the value $1$
problem only relies \emph{qualitatively} on the
asymptotic behavior of probabilities
(the probability to be in non-final states should be arbitrarily small)
the answer to the value $1$ problem
depends \emph{quantitatively} on the transition probabilities.

\begin{figure}[ht]
\begin{center}
\includegraphics[scale=1]{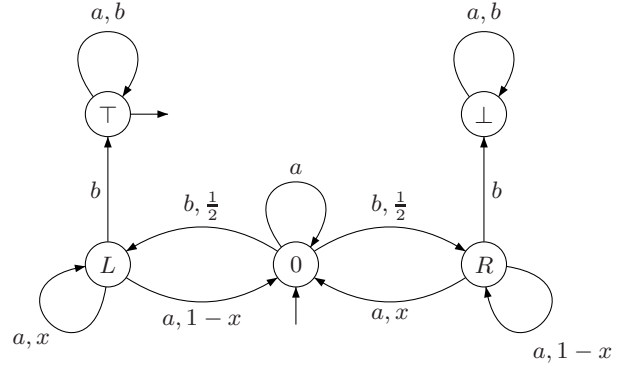}
\caption{\label{fig:x} This automaton has value $1$ if and only if $x>\frac{1}{2}$. }
\end{center}
\end{figure}

For instance, the automaton depicted on Fig.~\ref{fig:x}
has value $1$ if and only if $x > \frac{1}{2}$
and has value
less or equal than $\frac{1}{2}$ otherwise, see also~\cite{BBG08,GO10} for
similar results.
Note that in this example, the value is a discontinuous function of the transition probabilities. 
The input alphabet is $A =\set{a,b}$,
the initial state is the central
state $0$ and the unique final state is $\top$.
In order to maximize the probability to reach $\top$,
playing two $b$'s in a row is certainly not a good
option because from state $0$
this ensures to reach the non-accepting absorbing state
$\bot$ with probability at least $\frac{1}{2}$.
A smarter strategy consists in playing one $b$, then long sequences of $a$'s followed by
one letter $b$. If $x \leq \frac{1}{2}$, there is still no hope to have a word accepted
with probability strictly greater than $\frac{1}{2}$: starting from $0$,
and after a $b$ and a sequence of $a$'s, the probability to be in $R$ is greater
or equal than the probability to be in $L$, thus playing $ba^nb$ from state $0$
the probability to reach the sink $\bot$ is greater or equal than the
probability to reach the final state $\top$.
However, if $x > \frac{1}{2}$ then a simple calculation shows that the probability to accept $(ba^n)^n$ 
tends to $1$ as $n$ goes to infinity.

%The undecidability of the value $1$ problem comes from the necessity 
%to compare optimal parallel convergence rates
%in order to track down vanishing probabilities.
%Comparing two convergence rates may require to compare
%the decimals of the rates up to an arbitrary
%precision,
%which in turn can encode a Post correspondence problem,
%hence the undecidability.

\subsection{Undecidability in a very restricted case}

As a first step we refine the undecidability result:
we show that the value $1$ problem is undecidable even when restricted to probabilistic automata
having exactly one probabilistic transition.
For such automata, there exists exactly one state $s$ and one letter $a$ such that
$0 \leq  M_a(s,t)< 1$ for all $t$ and
the remaining transitions are deterministic:
for all triple $(s',a',t)\in S\times A \times S$
such that $(s',a')\neq (s,a)$ then
$M_a(s',t) \in \set{0,1}$.

The general idea is to simulate any
probabilistic automaton $\AA$ with a probabilistic automaton
$\BB$ which has only one probabilistic transition
and such that $\val{\AA} = 1$
if and only if $\val{\BB} = 1$.

As a first attempt, we define the automaton $\BB$ with a larger alphabet: 
whenever $\AA$ reads a letter $a$, then $\BB$ reads a sequence of actions $\widehat{a}$ corresponding to $a$, 
allowing a state-by-state simulation of $\AA$.
The unique probabilistic transition of $\BB$ is used to generate random bits for the simulation.
However, the automaton $\BB$ cannot check that the sequences of actions are well-formed
and allow for a faithful simulation.
Hence we modify the construction, such that to simulate the automaton $\AA$ on the input word $w$, 
the automaton $\BB$ now reads $(\widehat{w})^n$ for arbitrarily large $n$.
Each time $\BB$ reads a word $\widehat{w}$, it simulates $\AA$ on $w$ with a small yet positive probability
and ``delays'' the rest of the simulation, also with positive probability.
This delay process allows to run on parallel a deterministic automaton which checks
that the sequences of actions are well-formed, ensuring a faithful simulation.
The complete details can be found in the appendix (see also~\cite{FGO11}).

This undecidability result illustrates that even very restricted
classes of probabilistic automata may have an undecidable value $1$ problem.
In the next section, we introduce a non-trivial yet decidable subclass of probabilistic automata,
defined by the leaktight property.

\subsection{Informal description of the leaktight property}

One of the phenomena that makes tracking vanishing probabilities
difficult are \emph{leaks}.
A leak occurs in an automaton
when a sequence of words turns a set of states $C\subseteq Q$
into a recurrence class $C$ on the long run
but on the short run, some of the probability of the recurrence class
is ``leaking'' outside the class.

Such leaks occur in the automaton of Fig.~\ref{fig:x}
with the input sequence $(a^nb)\nNN$.
As $n$ grows large, the probability
to reach $\top$ and $\bot$ while reading the input word $a^nb$
vanishes: there are leaks from $L$ to $\top$ and symmetrically from $R$ to $\bot$.
As a consequence, the real asymptotic behavior is
complex and depends on the compared speeds of these leaks.

An automaton without leak is called a leaktight automaton.
In the next section we prove that the value $1$ problem
is decidable when restricted to the subclass
of leaktight automata.

The definition of a leaktight automaton
relies on two key notions, idempotent words
and word-recurrent states.

A finite word $u$ is \emph{idempotent} if reading once or twice the word
$u$ does not change qualitatively the transition probabilities:
 
\begin{definition}[Idempotent words]
A finite word $u\in A^*$ is idempotent
if
for every states $s,t\in Q$,
\[
u(s,t)>0 \iff (u\cdot u)(s,t)>0
%\prob{\AA}(s \xrightarrow{w} T
\enspace.
\]
\end{definition}

Idempotent words are everywhere:
every word, if iterated a large number of times,
becomes idempotent.

\begin{lemma}
\label{lem:idempotenteverywhere}
For every word $u\in A^*$,
the word $u^{|Q|!}$ is idempotent.
\end{lemma}

A finite word $u$ induces naturally a finite homogeneous Markov chain,
which splits the set of states into two classes: recurrent states
and transient states.
Intuitively, a state is transient if there is some non-zero probability to leave it forever,
and recurrent otherwise; equivalently from a recurrent state the probability to visit it again in the future is one.

\begin{definition}[Recurrent states]
Let $u\in A^*$ be a finite word.
A state $s$ is \emph{$u$-recurrent} if
it is recurrent in the finite Markov chain $\MM_u$ with states $Q$
and transitions probabilities $(u(s,t))_{s,t\in Q}$.
\end{definition}

Formally, $s$ in recurrent in $\MM_u$ if for all $t$ in $Q$, 
if there is a non-zero probability to reach $t$ from $s$,
then there is a non-zero probability to reach $s$ from $t$.

In the case of idempotent words,
recurrence of a state can be easily characterized:
\begin{lemma}\label{lem:idempotent}
Let $s$ be a state and $u$ be an idempotent word.
Then $s$ is $u$-recurrent if for every state $t$,
\[u(s,t)>0 \implies u(t,s) > 0\enspace.\]
\end{lemma}

The proof of this lemma follows from the observation that
since $u$ is idempotent, there is a non-zero probability to reach $t$ from $s$
if and only if $u(s,t) > 0$.

%\begin{IEEEproof}
%Suppose first that $s$ is $u$-recurrent,
%and let $t\in Q$ such that $u(s,t)>0$.
%Then $t$ is accessible from $s$ in $\MM_u$
%and by definition of recurrence,
%$s$ is accessible from $t$ in $\MM_u$
%thus there exists $k\in \NN$ such that
%$u^k(t,s)>0$.
%Since $u$ is idempotent,
%an easy induction shows that $\exists l>0, u^l(t,s)>0\implies u(t,s)>0$.
%
%Suppose now that for every $t\in Q$, $u(s,t)>0\implies u(t,s)>0$.
%Let $t$ be a state accessible from $s$,
%then there exists $k\in \NN$ such that
%$u^k(s,t)>0$.
%The same easy induction shows that $u(s,t)>0$
%thus by hypothesis $u(t,s)>0$ and $s$ is accessible from $t$ in $\MM_u$.
%Since this holds for every $t$ accessible from $s$,
%the state $s$ is recurrent in $\MM_u$.
%\end{IEEEproof}

The formal definition of a leak is as follows:

\begin{definition}[Leaks and leaktight automata]
\label{def:leak}
A leak from a state $r\in Q$ to a state $q\in Q$
is a sequence $(u_n)_{n\in \NN}$
of idempotent words such that:
\begin{enumerate}
\item for every $s,t\in Q$,
the sequence $(u_n(s,t))_{n\in\NN}$
converges to some value $u(s,t)$. 
We denote by $\mathcal{M}_u$ the Markov chain with states
$Q$ and transition probabilities $(u(s,t))_{s,t\in
Q}$,
\item
the state $r$ is recurrent in $\MM_u$,
\item
for all $n$ in $\NN$, $u_n(r,q)>0$,
\item
and $r$ is not reachable from $q$ in $\MM_u$.
\end{enumerate}

A probabilistic automaton is leaktight if it has no leak.
\end{definition}

The automaton depicted in Fig.~\ref{fig:x} is not leaktight
when $0 < x < 1$
because the sequence $(u_n)_{n\in\NN} = (a^nb)_{n\in\NN}$ is a leak
from $L$ to $\top$, and from $R$ to $\bot$.
The limit Markov chain $\mathcal{M}_u$
sends state $0$ to states $L$ and $R$ with probability $\frac{1}{2}$ each,
and all other states are absorbing (\textit{i.e} loop with probability $1$).
In particular, state $L$ is recurrent in $\mathcal{M}_u$,
for every $n$, $u_n(L, \top) > 0$ but
there is no transition from $\top$ to $L$ in $\mathcal{M}_u$.

Several examples of leaktight automata are given in Section~\ref{sec:ex}.

\section{\label{sec:dec}The value 1 problem is decidable for leaktight automata}

In this section we establish our main result:

\begin{theorem}
The value $1$ problem is decidable for leaktight automata.
\end{theorem}

\subsection{The Markov monoid algorithm}

Our decision algorithm for the value $1$ problem
computes iteratively
a set $\monoid$ of directed graphs
called limit-words.
Each limit-word is meant to
represent the asymptotic effect of
a sequence of input words,
and some particular limit-words
can witness that the automaton has value
$1$.

   \algsetup{indent=1em} 
   \begin{algorithm}[h!t]
\caption{
The Markov monoid algorithm.
}
\label{fig:thealgo}
     \algsetup{indent=2em, linenodelimiter=}
     \begin{algorithmic}[1] 
     \REQUIRE{A probabilistic automaton $\AA$.}
	\ENSURE{Decide whether $\AA$ has value $1$ or not.}
     
       \STATE{$\monoid \gets \{ \lima \mid a\in A\} \cup \{{\mathbf
       1}\}$.}
       \REPEAT
       \IF{there is $\limu,\limv\in \monoid$ such that $\limu\cdot \limv \notin \monoid$}
    \STATE{add $\limu\cdot \limv$ to $\monoid$}
 \ENDIF
 \IF{there is $\limu\in \monoid$ such that $\limu=\limu\cdot \limu$ and $\limu^\sharp \notin
 \monoid$}
 \STATE{add $\limu^\sharp$ to $\monoid$}
 \ENDIF
       \UNTIL{there is nothing to add}
 \IF{there is a value $1$ witness in  $\monoid$}
 
 \RETURN{\TRUE}
 \ELSE
\RETURN{\FALSE}
 \ENDIF
     \end{algorithmic} 
   \end{algorithm}

In the rest of the section, we explain the algorithm in details.

\begin{definition}[Limit-word]
A \emph{limit-word} is a map $\limu:Q^2\to\{0,1\}$
such that $\forall s\in Q, \exists t\in Q, \limu(s,t) = 1$.
\end{definition}

The condition expresses that our automata are complete:
whatever the input word, from any state $s$ there exists some state $t$
which is reached with positive probability.
A limit-word $\limu$ can be seen as a directed graph with no dead-end,
whose vertices are the states of the automaton $\AA$,
where there is an edge from $s$ to $t$ if $\limu(s,t) = 1$.

Initially, $\monoid$ only contains those limit-words $\lima$
that are induced by input letters $a\in A$,
where the limit-word $\lima$ is defined by:
\[
\forall s,t\in Q, (\lima(s,t)=1 \iff a(s,t)>0)\enspace.
\]
plus the identity limit-word $\mathbf 1$
defined by $({\mathbf 1}(s,t)=1) \iff (s=t)$,
which represents the constant sequence of the empty word.

The algorithm repeatedly adds new limit-words to $\monoid$.
There are two ways for that:
concatenating two limit-words in $\monoid$
or iterating an idempotent limit-word in $\monoid$.

\paragraph*{Concatenation of two limit-words}
The \emph{concatenation} of two limit-words $\limu$
and $\limv$ is the limit-word $\limu \cdot \limv$ such that:
\[
(\limu \cdot \limv)(s,t) = 1 \iff \exists q\in Q, \limu(s,q)=1 \text{ and }
\limv(q,t)=1 \enspace.
\]
In other words, concatenation coincides with the multiplication
of matrices with coefficients in the boolean semiring
$(\{0,1\},\vee,\wedge)$.
The concatenation of two limit-words intuitively corresponds to
the concatenation of two sequences $(u_n)_{n\in\NN}$ and $(v_n)_{n\in\NN}$ of input
words into the sequence $(u_n\cdot v_n)_{n\in\NN}$.
Note that the identity limit-word $\mathbf 1$ is neutral for the concatenation.

\paragraph*{Iteration of an idempotent limit-word}
Intuitively, if a limit-word $\limu$ represents a
sequence $(u_n)_{n\in\NN}$ then its iteration $\limu^\sharp$
represents the sequence $\left(u_n^{f(n)}\right)_{n\in\NN}$
for an arbitrarily large increasing function $f:\NN\to\NN$.

The \emph{iteration} $\limu^\sharp$ of a limit-word $\limu$ is only defined
when $\limu$ is idempotent \textit{i.e} when $\limu \cdot \limu = \limu$.
It relies on the notion of $\limu$-recurrent state.
\begin{definition}[$\limu$-recurrence]
Let $\limu$ be an idempotent limit-word.
A state $s$ is $\limu$-recurrent if for
every state $t$,
\[
\limu(s,t)=1 \implies \limu(t,s)=1\enspace.
\]
\end{definition}

%Lemma~\ref{lem:idempotent} shows the link between the notions of recurrent
%states for limit-words on one hand and for Markov chains induced by finite
%words on the other hand.

The \emph{iterated limit-word} $\limu^\sharp$ removes from $\limu$
any edge that does not lead to a recurrent state:
\[
\limu^\sharp(s,t)=1 \iff \limu(s,t)=1 \text{ and } t \text{ is }
\limu\text{-recurrent}
\enspace.
\]

\subsection{The Markov monoid and value $1$ witnesses}

The set $\monoid$ of limit-words computed by the Markov monoid algorithm is called the Markov monoid.

\begin{definition}[Markov monoid]
The Markov monoid is the smallest set of limit-words containing
the set $\{\lima\mid a\in A\}$ of limit-words induced by letters, the identity
limit-word $\mathbf 1$, and closed under concatenation and iteration.
\end{definition}

Two key properties,
\emph{consistency} and \emph{completeness},
ensure that the limit-words of the Markov monoid
reflect exactly every 
possible asymptotic effect
of a sequence of input words.

Consistency ensures that every limit-word in $\monoid$ abstracts
the asymptotic effect of an input sequence.

 \begin{definition}[Consistency]
\label{def:consistency}
A set of limit-words
$\monoid\subseteq \{0,1\}^{Q^2}$ is
\emph{consistent}
with a probabilistic automaton $\AA$
if
for each limit-word $\limu\in\monoid$, there exists a sequence of input words
$(u_n)_{n\in\NN}$
such that for every states $s,t\in Q$
the sequence $(u_n(s,t))_{n\in\NN}$ converges and
\begin{equation}
\label{eq:consistency}
\limu(s,t)=1\iff \lim_n u_n(s,t)>0\enspace.
\end{equation}
\end{definition}

Conversely, completeness ensures that every input sequence
reifies one of the limit-words.

\begin{definition}[Completeness]
\label{def:complete}
A set of limit-words
$\monoid\subseteq \{0,1\}^{Q^2}$ is
\emph{complete}
for a probabilistic automaton $\AA$
if
for each sequence of input words
$(u_n)_{n\in\NN}$, there exists $\limu\in\monoid$
such that for every states $s,t\in Q$:
\begin{equation}
\label{eq:completeness}
\limsup_n u_n(s,t)=0 \implies \limu(s,t)=0\enspace.
\end{equation} 
\end{definition}

A limit-word may witness that the automaton has value $1$.

\begin{definition}[Value $1$ witnesses]
\label{def:value1witnesses}
Let $\AA$ be a probabilistic automaton.
A \emph{value $1$ witness} is a limit-word $\limu$ such that
for every state $s\in Q$,
\begin{equation}
\label{eq:witness}
\limu(i,s)=1 \implies s\in F
\enspace.
\end{equation}
\end{definition}

Thanks to value $1$ witnesses,
the answer to the value $1$ problem can be read in a
consistent and complete set of limit-words:

\begin{lemma}[A criterion for value $1$]
\label{lem:criterion}
Let $\AA$ be a probabilistic automaton
and $\monoid\subseteq \{0,1\}^{Q^2}$ be a set of limit-words.
Suppose that $\monoid$ is
consistent with $\AA$ and complete for $\AA$.
Then $\AA$ has value $1$ if and only if $\monoid$ contains a value $1$
witness.
\end{lemma}

In order to illustrate the interplay between limit-words of the Markov monoid
and sequences of input words, we give a detailed proof of
Lemma~\ref{lem:criterion}.

\begin{IEEEproof}
Assume first that $\AA$ has value $1$.
By definition, there exists a sequence $(u_n)_{n\in\NN}$ of input words
such that $\prob{\AA}(u_n) \tendsto{n} 1$.
As a consequence,
$\sum_{f\in F} u_n(i,f) = \prob{\AA}(u_n)\tendsto{n}  1$.
Since for all $n\in \NN$, we have $\sum_{q\in Q} u_n(i,q)=1$,
then for all $s'\notin F$, $u_n(i,s')\tendsto{n} 0$.
Since $\monoid$ is complete,
there exists a limit-word $\limu$ such
that~\eqref{eq:completeness} holds.
Then $\limu$ is a value $1$ witness:
for every $s\in Q$ such that $\limu(i,s)=1$,
equation~\eqref{eq:completeness} implies
$\limsup_n u_n(i,s)>0$,
hence $s\in F$.

Conversely, assume now that $\monoid$ contains a value $1$ witness $\limu$.
Since $\monoid$ is consistent,
there exists a sequence $(u_n)\nNN$ such that~\eqref{eq:consistency} holds.
It follows from~\eqref{eq:consistency} and~\eqref{eq:witness},
that for all $s\not\in F$, we have $u_n(i,s)\tendsto{n} 0$. 
Thus 
$\prob{\AA}(u_n) = \sum_{f\in F} u_n(i,f)\tendsto{n} 1$ 
and $\AA$ has value $1$.
\end{IEEEproof}

The following theorem proves that the Markov monoid of a leaktight automaton
is consistent and complete,
thus according to Lemma~\ref{lem:criterion} it
can be used to decide the value $1$ problem.

\begin{theorem}
\label{theo:good}
The Markov monoid associated with an automaton $\AA$ is consistent.
Moreover if $\AA$ is leaktight then the Markov monoid is complete. 
\end{theorem}

The proof of the second part of this theorem relies on a subtle algebraic argument
based on the existence of factorization forests of bounded
height~\cite{simonforest}.
The same kind of argument was used by Simon to prove the decidability of the
boundedness problem for distance automata~\cite{simontropical}.
%Note that the boundedness problem and the value $1$
%problem for leaktight automata seem hardly reducible to each other,
%because, in a nutshell, the probabilistic monoid $(\RR,+,\cdot)$
%and the tropical semi-ring $(\NN\cup \{ \infty\},\min,+)$
%are very different. See~\cite{GO10} for more details about the links
%and differences between probabilistic and non-deterministic automata.

We postpone the proof of completeness to the next section,
where a slightly more general result is established;
for now we show that the Markov monoid is consistent.

\begin{lemma}[Consistency]\label{lem:consistency}
Let $\monoid\subseteq \{0,1\}^{Q^2}$ be a set of limit-words.
Suppose that $\monoid$ is consistent.
Then for every $\limu,\limv\in \monoid$ the set $\monoid \cup\{\limu\cdot
\limv\}$ is consistent. If moreover $\limu$ is idempotent then $\monoid
\cup\{\limu^\sharp\}$ is consistent as well.
\end{lemma}

The proof uses the notion of reification.

\begin{definition}
A sequence $(u_n)\nNN$ of input words reifies a limit-word $\limu$
if for every states $s,t$ the sequence $(u_n(s,t))_{n\in\NN}$ converges and
\begin{equation}
\limu(s,t)=1\iff \lim_n u_n(s,t)>0\enspace.
\end{equation}
In particular, a set of limit-words $\monoid$ is consistent for $\AA$ if each
limit-word in $\monoid$ is reified by some sequence of input words.
\end{definition}

\begin{IEEEproof}
Let $\limu,\limv\in \monoid$.
We build a sequence $(w_n)\nNN$ which reifies $\limu\cdot\limv$.
By induction hypothesis on $\limu$ and
$\limv$, there exists $(u_{n})_n$ and $(v_{n})_n$
which reify $\limu$ and $\limv$ respectively.
Let $w_n = u_{n} \cdot v_{n}$.
Then $(w_n)\nNN$ reifies $\limu\cdot \limv$, because
\[
w_n(s,r) = \sum_{t \in Q} u_n(s,t) \cdot v_n(t,r)\enspace
\]
and by definition of the concatenation of two limit-words.

Suppose now that $\limu$ is idempotent,
we build a sequence $(z_n)\nNN$ which reifies $\limu^\sharp$.
By induction hypothesis, there exists a sequence  $(u_n)\nNN$ which reifies
$\limu$.
For every states $s,t$ we denote by $u(s,t)$ the value $\lim_n u_n(s,t)$.
% Since $[0,1]^{Q\times Q}$ is compact,
%we can suppose (by considering subsequences if needed) that for every $s,t\in
% Q$, $u_n(s,t)$ converges to some value $u(s,t)$.
Since $\limu$ is idempotent, the Markov chain $\MM_u$ with state space $Q$
and transition probabilities $(u(s,t))_{s,t\in Q}$
is 1-periodic thus aperiodic.
%, the onidempotent as well: for all states $s,t$
%there is a non-zero probability to reach $t$ from $s$ in $\MM_u$ if and only if
% $u(s,t) > 0$.
According to standard results about finite Markov chains,
the sequence of matrices
$(u^k)_{k\in\NN}$ has a limit $z\in [0,1]^{Q\times Q}$
such that transient states of $\mathcal{M}_u$
have no incoming edges in $z$.
This implies:
\begin{equation}
\label{eq:trans}
\forall s,t \in Q,\ (z(s,t)>0 \implies t\text{ is $z$-recurrent})\ .
\end{equation} 
Since $(u_n)\nNN$ converges to $u$
and by continuity of the matrix product,
for every $k\in\NN$ the sequence of matrices $(u_n^k)_{n\in\NN}$ converges to $u^k$.
It follows that there exists $\phi(k)\in\NN$ such that $||u^k - u_{\phi(k)}^k ||_\infty \leq \frac{1}{k}$.
As a consequence the sequence of matrices $(z_n)_{n\in\NN} =
(u_{\phi(n)}^n)_{n\in\NN}$ converges to $z$.

Now we prove that $(z_n)\nNN$ reifies
$\limu^\sharp$ because,
\begin{align*}
\limu^\sharp(s,t)=1
& \iff t \text{ is } \limu\text{-recurrent and } \limu(s,t)=1\\
& \iff t \text{ is } u\text{-recurrent and } u(s,t)>0\\
& \iff t \text{ is } z\text{-recurrent and } z(s,t)>0\\
& \iff z(s,t)>0\\
& \iff \lim_n z_n(s,t)>0
\enspace,
%) \lim_n u_n(s,t)>0 \iff \lim_k (u_{\phi(k)}^k)(s,t)>0 \iff \lim_n
%z_n(s,t)>0.
\end{align*}
where the first equivalence is by definition of the iteration,
the second holds because $(u_n)\nNN$ reifies $\limu$,
the third because the iterated Markov chain induced by $z=\lim_k u^k$
has the same recurrent states than the Markov chain $\mathcal{M}_u$,
the fourth holds by~\eqref{eq:trans},
and the fifth by definition of $z$.
\end{IEEEproof}

\subsection{\label{subsec:algo}Correctness of the Markov monoid algorithm}

\begin{proposition}\label{prop:correctness}
The Markov monoid algorithm solves the value $1$ problem
for leaktight automata.
\end{proposition}

\begin{IEEEproof}
Termination of the Markov monoid algorithm
is straightforward because each iteration adds a new element
in $\monoid$ and there are at most $2^{|Q|^2}$
elements in $\monoid$.

The correctness is a corollary of Theorem~\ref{theo:good}:
since the Markov monoid is consistent and complete then according to
Lemma~\ref{lem:criterion},
$\AA$ has value $1$ if and only if $\monoid$ contains a value $1$ witness,
if and only if the Markov monoid algorithm outputs ``true''.
\end{IEEEproof}

%In the case when the Markov monoid algorithm outputs ``true'',
In case the Markov monoid algorithm outputs ``true'',
then for sure the input automaton has value
$1$. This positive result holds for every automaton,
leaktight or not.

\begin{proposition}
\label{prop:yes}
If the Markov monoid algorithm outputs ``true'', the input probabilistic
automaton has value $1$.
\end{proposition}

\begin{IEEEproof}
According to Theorem~\ref{theo:good},
the Markov monoid is consistent.
If it contains a value $1$ witness,
then according to the second part of the proof of
Lemma~\ref{lem:criterion}, $\AA$ has value $1$.
\end{IEEEproof}

In case the Markov monoid algorithm outputs ``false''
and the automaton is leaktight
then the value of the automaton
can be bounded from above:
% has value strictly less than one.

\begin{theorem}
\label{theo:no}
Let $\AA$ be a probabilistic automaton whose minimal
non-zero transition probability is denoted $\pmin$.
If the Markov monoid algorithm outputs ``false''
and if moreover $\AA$
is leaktight,
then $\val{\AA}\leq 1-\dabound$,
with $\numjclasses = \explicitnumjclasses$.
\end{theorem}

The proof of this theorem is postponed to the next section,
because it relies on the notion of extended Markov monoid,
it is actually a direct corollary of the lower bound lemma
presented in Section~\ref{sec:correct}.

In case the Markov monoid algorithm outputs ``false'', 
one surely wishes to know whether the input automaton
is leaktight or not.
Fortunately, the leaktight property is decidable,
this is the subject of the next section.

\subsection{\label{subsec:algocompl}Complexity of the Markov monoid algorithm}

\begin{proposition}\label{prop:pspacecomplete}
The value $1$ problem for leaktight automata is $\PSPACE$-complete.
\end{proposition}

The Markov monoid algorithm 
terminates in less than $2^{|Q|^2}$ iterations,
since each iteration adds a new limit-word in the monoid
and there are less than $2^{|Q|^2}$ different limit-words.

This $\EXPTIME$ upper bound can be actually improved to $\PSPACE$.
For that we use the same arguments that Kirsten
used to prove that limitedness of desert automata
can be decided in $\PSPACE$~\cite{Kirsten05}.

A way to improve the complexity from $\EXPTIME$ to $\PSPACE$
is to avoid the explicit computation of the Markov monoid
and to look for value $1$ witnesses in a
non-deterministic way.
The algorithm guesses non-deterministically the value $1$ witness $\limu$
and its
decomposition by the product and iteration operations.
The algorithm computes a $\sharp$-expression,
\textit{i.e} a finite tree
with concatenation nodes of arbitrary degree on even levels
and iteration nodes of degree $1$ on odd levels
and labelled consistently by limit-words.
The depth of this tree is at most twice the $\sharp$-height (the number of
nested applications of the iteration operation) plus $1$.
The root of the $\sharp$-expression is labelled by $\limu$
and the expression is computed
non-deterministically from the root in a depth-first way.

For desert automata, the key observation made by Kirsten is that the
$\sharp$-height is at most $|Q|$.
The adaptation of Kirsten's proof to probabilistic automata relies on the
two following lemmata:
\begin{lemma}\label{lem:rec_class_inclu}
Let $\limu$ and $\limv$ be two idempotent limit-words. Assume $\limu \leq_\JJ \limv$,
then there are less recurrence classes in $\limu$ than in
$\limv$.
\end{lemma}

\begin{lemma}\label{lem:rec_class_inclu_strict}
Let $\limu$ be an idempotent limit-word. The set of recurrence classes of
$\limu$ is included in the set of recurrence classes of $\limu^\sharp$.
Moreover if $\limu \neq \limu^\sharp$ this inclusion is strict.
\end{lemma}

Since the number of recurrence classes in a limit-word is bounded by $|Q|$,
and if we require the iteration operation to be applied only
to unstable idempotent,
the $\sharp$-height of a $\sharp$-expression is bounded
by $|Q|$ thus the depth of the expression is bounded by $2|Q|+1$.

%Relying on the two remarks, we can show that the Markov monoid algorithm uses
%polynomial space~\cite{Kirsten05}. 
Consequently,
the value $1$ problem can be decided in $\PSPACE$:
to guess the value $1$ witness,
the non-deterministic algorithm needs to store at most $2|Q|+1$ limit-words
which can be done in space $\mathcal{O}(|Q|^2)$.
Savitch's theorem implies that the deterministic complexity is $\PSPACE$
as well.

%Relying on those two remarks,
%we can show that the Markov monoid algorithm uses polynomial
%space~\cite{Kirsten05}. 

This $\PSPACE$-upperbound on the complexity is tight.
The value $1$ problem is known to be $\PSPACE$-complete
when restricted to $\sharp$-acyclic automata~\cite{GO10}.
The same reduction to the $\PSPACE$-complete problem of
intersection of deterministic automata can be used
to prove completeness of the value $1$ problem for leaktight automata,
relying on the facts that deterministic automata are leaktight
(Proposition~\ref{prop:containment}) and the class of leaktight
automata is closed under parallel composition (Proposition~\ref{prop:stable}).
The completeness result is also a corollary of
Proposition~\ref{prop:containment}:
since $\sharp$-acyclic automata are a subclass of leaktight automata,
the decision problem is \emph{a fortiori} complete for leaktight automata.

\section{\label{sec:leak}Deciding whether an automaton is leaktight}

At first sight,
the decidability of the leaktight property is not obvious:
to check the existence of a leak one would need
to scan the uncountable set of all possible sequences of input words.
Still:

\begin{theorem}
The leaktight property is decidable in polynomial space.
\end{theorem}

  \algsetup{indent=1em} 
   \begin{algorithm}[h!t]
\caption{
\label{algo:extended}
The leak-finder algorithm.
}
\label{fig:theleakfinder}
     \algsetup{indent=2em, linenodelimiter=}
     \begin{algorithmic}[1] 
     \REQUIRE{A probabilistic automaton $\AA$.}
	\ENSURE{Decide whether $\AA$ is leaktight or not.}
     
       \STATE{$\monoidext \gets \{ (\lima,\lima) \mid a\in A\} \cup \set{(\mathbf 1,\mathbf 1)}$.}
       \REPEAT
       \IF{there is $(\limu,\limu_+),(\limv,\limv_+)\in \monoidext$ such that $(\limu\cdot
       \limu,\limv_+\cdot \limv_+)\not \in \monoidext$}
    \STATE{add $(\limu\cdot \limv,\limu_+\cdot \limv_+)$ to $\monoidext$}
 \ENDIF
 \IF{there is $(\limu,\limu_+)\in \monoidext$ such that $\limu=\limu\cdot \limu$ and $\limu_+=\limu_+\cdot
 \limu_+$ and $(\limu^\sharp,\limu_+)\not \in \monoidext$}
 \STATE{add $(\limu^\sharp,\limu_+)$ to $\monoidext$}
 \ENDIF
       \UNTIL{there is nothing to add}
 \IF{there is a leak witness in $\monoidext$}
 \RETURN{\FALSE}
 \ELSE
\RETURN{\TRUE}
 \ENDIF
     \end{algorithmic} 
   \end{algorithm} 

The \emph{leak-finder algorithm} deciding the leaktight property is very
similar to the Markov monoid algorithm, except for two differences.
First, the algorithm keeps track of those edges that are deleted by
successive iteration operations. For that purpose, the algorithm
stores together with each limit-word $\limu$ another limit-word $\limu_+$
to keep track of strictly positive transition probabilities.
Second, the algorithm looks for \emph{leak witnesses}.

\begin{definition}[Extended limit-word]
An extended limit-word is a pair of limit-words.
The set of extended limit-words computed by the leak-finder algorithm is called the \emph{extended Markov monoid}.
\end{definition}

The extended Markov monoid is indeed a monoid equipped with the component-wise concatenation operation:
\[
(\limu,\limu_+)\cdot(\limv,\limv_+)= (\limu\cdot \limv,\limu_+\cdot
\limv_+)\enspace,
\]
It follows that an extended limit-word $(\limu,\limu_+)$ is idempotent if both $\limu$ and $\limu_+$ are idempotent.

\begin{definition}[Leak witness]
\label{eq:leakwitness}
An extended limit-word $(\limu,\limu_+)$ is a \emph{leak witness} if it is idempotent
and there exists $r,q\in Q$ such that:
\begin{enumerate}
  \item $r$ is $\limu$-recurrent,
  \item $\limu_+(r,q)=1$,
  \item $\limu(q,r)=0$.
\end{enumerate}
%
%If the extended Markov monoid of an automaton $\AA$
%contains a leak witness then $\AA$ has a leak.
\end{definition}

The correctness of the leak-finder algorithm is a consequence of:
\begin{theorem}
\label{theo:leakwitness}
An automaton $\AA$ is leaktight if and only if its
extended Markov monoid does not contain a leak witness.
\end{theorem}

The proof can be found in the appendix.
Although we chose to present Theorem~\ref{theo:good} and
Theorem~\ref{theo:leakwitness} separately,
their proofs are tightly linked.

As a consequence, the leaktight property is qualitative:
it does not depend on the exact value of transition probabilities
but only on their positivity.

\section{\label{sec:correct}The lower bound lemma}

The lower bound lemma is the key to both our decidability result
(via Theorem~\ref{theo:no})
and the characterization of leaktight automata
(Theorem~\ref{theo:leakwitness}).

\begin{lemma}[Lower bound lemma]\label{lem:lowerbound}
Let $\AA$ be a probabilistic automaton whose
extended Markov monoid contains no leak witness.
Let $\pmin$ the smallest non-zero transition probability of $\AA$.
Then for every word $u \in A^*$, there exists a pair
$(\limu,\limu_+)$  in the extended Markov monoid
such that, for all states $s,t
\in Q$,
\begin{align}
\label{eq:lb1}
\limu_+(s,t)=1 &\iff u(s,t)>0\enspace,\\
\label{eq:lb2}
\limu(s,t) = 1 &\implies  u(s,t) \geq \dabound \enspace,
\end{align}
where $\numjclasses = 2^{2|Q|^2}$.
\end{lemma}

To prove Lemma~\ref{lem:lowerbound}, we rely on the notion of Ramseyan
factorization trees and decomposition trees introduced by
Simon~\cite{simonforest,simontropical}.

\begin{definition}
\label{def:ramsey}
Let $A$ be a finite alphabet, $(M,\cdot,1)$ a monoid and $\phi: A^* \to M$ a
morphism. A \emph{Ramseyan factorization tree} of a word $u \in A^+$ for $\phi$
is a finite unranked ordered tree, whose nodes are labelled by pairs
$(w,\phi(w))$ where $w$ is a word in $A^+$ and such that:
\begin{enumerate}
	\item[(i)] the root is labelled by $(u,\phi(u))$,
	\item[(ii)] every internal node with two children labelled by 
$(u_1,\phi(u_1))$ and $(u_2,\phi(u_2))$ is labelled by 
$(u_1 \cdot u_2,\phi(u_1\cdot u_2))$,
	\item[(iii)] leaves are labelled by pairs $(a,\phi(a))$ with $a\in A$,
	\item[(iv)] if an internal node $t$ has three or more children $t_1,\ldots,
	t_n$ labelled by $(u_1,\phi(u_1)),\ldots,(u_n,\phi(u_n))$,
	then there exists $\lime\in M$ such that $\lime$ is idempotent and
	$\lime=\phi(u_1)=\phi(u_2)=\ldots=\phi(u_n)$.
	In this case $t$ is labelled by $(u_1\cdots u_n,\lime)$.
%	an idempotent  then
%	the labels of all its children are mapped by $\phi$ to $e \in E(M)$ (in
%	particular, we also have that $e$ labels the node).
\end{enumerate}
Internal nodes with one or two children are \emph{concatenation nodes}, 
the other internal nodes are \emph{iteration nodes}.
\end{definition}

Not surprisingly, every word $u \in A^+$ can be factorized in a Ramseyan
factorization tree, using only concatenation nodes:
any \emph{binary} tree whose leaves are labelled from left to right by the
letters of $u$ and whose internal nodes are labelled consistently
is a Ramseyan factorization tree.
Notice that if $u$ has length $n$ then such a
tree has height $\log_2(n)$, with the convention that the height of a
leaf is $0$.
As a consequence, with this na\"ive factorization of $u$,
the longer the word $u$, the deeper its factorization tree.

The following powerful result of Simon states that
every word can be factorized with a
Ramseyan factorization tree whose depth
is bounded independently of the length of the word:

\begin{theorem}[\cite{simonforest,chalopinLeung,factocolcombet}]\label{theo:simon}
Let $\AA$ be a probabilistic automaton whose
extended Markov monoid contains no leak witness.
Every word $u \in A^+$ has a Ramseyan factorization tree of height at most 
$3 \cdot |M|$.
\end{theorem}

In~\cite{simontropical},
Simon used the tropical semiring $(\NN\cup\{\infty\},\min,+)$
to prove the decidability of the boundedness problem
for distance automata.
Similarly to the Markov monoid, the tropical semiring is equipped with an
iteration operation $\sharp$.
Following the proof scheme of Simon, we introduce the notion of decomposition
tree relatively to a monoid $M$ equipped with an iteration operation $\sharp$.

\begin{definition}
\label{def:decomp}
Let $A$ be a finite alphabet, $(M,\cdot,1)$ a monoid equipped with a
function $\sharp$ that maps every idempotent $\lime\in M$
to another idempotent element $\lime^\sharp\in M$
and $\phi: A^* \to M$ a morphism. 
A \emph{decomposition tree} of a word $u \in A^+$ is a finite unranked ordered
tree, whose nodes have labels in $(A^+,M)$ and such that:
\begin{itemize}
	\item[i)] the root is labelled by $(u,\limu)$, for some $\limu \in M$,
	\item[ii)] every internal node with two children labelled by 
$(u_1,\limu_1)$ and $(u_2,\limu_2)$ is labelled by 
$(u_1 \cdot u_2,\limu_1 \cdot \limu_2)$,
	\item[iii)] every leaf is labelled by $(a,\lima)$ where $a$ is a letter,
	\item[iv)] for every internal node with three or more children, 
there exists $\lime \in M$ such that $\lime$ is idempotent and
the node is labelled by $(u_1 \ldots u_n,\lime^\sharp)$
and its children are labelled by $(u_1,\lime),\ldots,(u_n,\lime)$.
\end{itemize}
Internal nodes with one or two children are \emph{concatenation} nodes, 
the other internal nodes are \emph{iteration} nodes.

An iteration node labelled by $(u,\lime)$
is \emph{discontinuous} if $\lime^\sharp\neq \lime$.
The \emph{span} of a decomposition tree is the maximal length of a path that
contains no discontinuous iteration node.
\end{definition}

Remark that decomposition and factorization trees are closely related:
\begin{lemma}\label{lem:decomp}
A Ramseyan factorization tree is a decomposition tree if and only if
it contains no discontinuous iteration nodes.
\end{lemma}
\begin{IEEEproof}
The definitions~\ref{def:ramsey} and~\ref{def:decomp}
are similar except for condition iv).
If there are no discontinuous nodes then $\lime=\lime^\sharp$
in iv) of Definition~\ref{def:decomp}.
\end{IEEEproof}
%Note that by definition, whenever a node is labelled by $(u,\limuu)$, then $w$
%reifies $u$.

%In a Ramsey factorization forest, as well as in a decomposition tree,
The following theorem is adapted from \cite[Lemma 10]{simontropical}
and is a direct corollary of Theorem~\ref{theo:simon}.
 
\begin{theorem}\label{theo:decomposition}
Let $A$ be a finite alphabet, $(M,\cdot,1)$ a monoid equipped with a
function $\sharp$ that maps every idempotent $\lime\in M$
to another idempotent element $\lime^\sharp\in M$
and $\phi: A^* \to M$ a morphism.
Every word $u \in A^+$ has a decomposition tree whose span is less than 
$3 \cdot |M|$.
\end{theorem}

To obtain the lower bound lemma,
we need to bound the depth of a decomposition tree;
now that the span is bounded thanks to Theorem~\ref{theo:decomposition},
we need to bound the number of discontinuous iteration nodes.
Simon and Leung noticed that this number is actually
bounded by the number of $\mathcal{J}$-classes in the monoid.
The notion of $\JJ$-class of a monoid $M$
is a classical notion in
semigroup theory, derived from one of the four Green's relations
called the $\JJ$-preorder: a $\JJ$-class is an equivalence class
for this preorder (for details about Green's relations,
see~\cite{howie,lallement}).
The $\JJ$-preorder between elements of a monoid $M$ is defined as follows:
\[
\forall a,b\in M, a\leq_\JJ b \text{ if } a\in MbM \enspace,
\]
where $MbM$ denotes the set $\{ubv\mid u,v \in M\}$.

The number of discontinuous nodes along a path in a decomposition tree can be
bounded using the following result, adapted from~\cite[Lemma 3]{simontropical}.

\begin{lemma}
\label{lem:discont}
Let $A$ be a finite alphabet, and $M$ a monoid equipped with a
function $\sharp$ that maps every idempotent $e\in M$
to another idempotent element $e^\sharp\in M$.
Suppose moreover that for every idempotent $e \in M$,
\begin{equation}
\label{eq:esharp}
e^\sharp \cdot e =
e^\sharp = e \cdot e^\sharp\enspace.
\end{equation}

Then for every idempotent element $e\in M$,
either $e^\sharp=e$ or $e^\sharp<_\JJ e$.

As a consequence, the number of discontinuous nodes along
a path in a decomposition tree is at most $\numjclasses$,
where $\numjclasses$ is the number of $\mathcal{J}$-classes of the monoid.
\end{lemma}

%\begin{IEEEproof}
%We prove the first part of the lemma.
%Equation~\eqref{eq:esharp}
%implies that $e^\sharp = e^\sharp e e^\sharp$ thus $e^\sharp \leq_\JJ e$.
%Now, we suppose that $e \leq_\JJ e^\sharp$ and prove that $e=e^\sharp$.
%Since $M$ is finite, we have $e \DD e^\sharp$.
%Since $e \cdot e^\sharp = e^\sharp$, it follows $e \RRR e^\sharp$.
%By a dual argument, we have $e \LL e^\sharp$; hence $e \HH e^\sharp$.
%Both $e$ and $e^\sharp$ and idempotents, so according to Green's theorem
%(see \textit{e.g}~\cite{howie}, Theorem 2.2.5.)
%$e = e^\sharp$.
%
%The second part of the lemma is an immediate consequence, 
%since the sequence of elements of the monoid labelling a branch of a decomposition
%tree, starting from the root,
%is non-decreasing for the $\JJ$-order and strictly
%increasing on discontinuous nodes.
%Hugo: I put back the end of the proof, if we 
%Indeed, if an internal node has two children labelled by 
%$(u_1,\limu_1)$ and $(u_2,\limu_2)$ then by definition of a decomposition tree
%the node is labelled by $(u_1 \cdot u_2,\limu_1 \cdot \limu_2)$
%and by definition of the $\JJ$-order,
%$\limu_1 \cdot \limu_2 \leq_\JJ \limu_1$ and $\limu_1 \cdot \limu_2 \leq_\JJ \limu_2$.
%For continuous nodes, the same idempotent labels both the internal node and its
%children. For discontinuous nodes, the father node is labelled by some idempotent
%$\lime^\sharp$ and the children by $\lime \neq \lime^\sharp$, and we conclude
%with supra.
%\end{IEEEproof}

Now we are ready to complete the proof of the lower bound lemma.

\begin{IEEEproof}[Proof of Lemma~\ref{lem:lowerbound}]
% We are now able to prove Lemma~\ref{lem_bound},
% which states that there exists a positive rational number $\eta$,
% which depends only on $\AA$, with the following property:
% for all words $w \in A^*$, there exists a $\sharp$-sentence $u$ such that,
% for all states $s,t \in Q$,
% $$\treps{u}(s,t) = 1 \Rightarrow \prob{\AA}(s \xrightarrow{w} t) \geq \eta.$$
% 
Let $M$ be the extended Markov monoid $\monoid_+$ associated with $\AA$
and equipped with the concatenation operation:
\[
(\limu,\limu_+)\cdot(\limv,\limv_+)= (\limu\cdot \limv,\limu_+\cdot
\limv_+)\enspace,
\]
and for idempotent pairs the iteration operation:
\[
(\limu,\limu_+)^\sharp = (\limu^\sharp,\limu_+)\enspace.
\]

Let $w \in A^+$. 
(The case of the empty word is easily settled, considering the extended limit-word $(\mathbf{1},\mathbf{1})$.)
We apply Theorem~\ref{theo:decomposition}
to the word $w$, the extended Markov monoid $M=\monoid_+$
and the morphism $\phi: A \to M$
defined by $\phi(a) = (\lima,\lima)$.
According to Theorem~\ref{theo:decomposition},
$w$ has a decomposition tree $T$ of span less than $3 \cdot |\monoid_+|$,
whose root is labelled by $(w,(\limw,\limw_+))$ for some extended limit-word
$(\limw,\limw_+) \in \monoid_+$.

According to the second part of Lemma~\eqref{lem:discont},
and since there are less $\JJ$-classes than there are elements in the monoid
$\monoid_+$,
\begin{equation}
\label{eq:depth}
\text{the depth of $T$ is at most $3 \cdot |\monoid_+|^2$.}
\end{equation}

\medskip

To complete the proof of Lemma~\ref{lem:lowerbound},
we prove that for every node $t$ labelled $(u,(\limu,\limu_+))$ of depth $h$
in the decomposition tree and for all states $s,t \in Q$,
\begin{align}
\label{eq:induc1}
\limu_+(s,t)=1 &\iff u(s,t)>0\enspace,\\
\label{eq:induc2}
\limu(s,t) = 1 &\implies u(s,t) \geq \pmin^{2^{h}} \enspace.
\end{align}

We prove~\eqref{eq:induc1} and~\eqref{eq:induc2} by induction on $h$.

If $h = 0$ then the node is a leaf,
hence $u$ is a letter $a$ and $\limu = \limu_+ = \lima$.
Then~\eqref{eq:induc1} holds by definition of $\lima$
and~\eqref{eq:induc2} holds by definition of $\pmin$.

For the induction, there are two cases.

{\bf First case, $t$ is a concatenation node} labelled
by $(u,(\limu,\limu_+))$ with two sons labelled
by $(u_1,(\limu_1,\limu_{+,1}))$
and $(u_2,(\limu_2,\limu_{+,2}))$.
We first prove that~\eqref{eq:induc1} holds.
Let $s,t\in Q$ such that $\limu_+(s,t)=1$.
By definition of a decomposition tree,
$\limu_+=\limu_{+,1}\cdot \limu_{+,2}$.
Since $\limu_+(s,t)=1$ then by definition
of the concatenation there exists $q\in Q$ such that
$\limu_{+,1}(s,q)=1$ and $\limu_{+,2}(q,t)=1$.
By induction hypothesis we have $u_1(s,q) \cdot u_2(q,t)>0$;
and since $u=u_1\cdot u_2$ then $u(s,t) \geq u_1(s,q) \cdot u_2(q,t)$,
which proves the direct implication of~\eqref{eq:induc1}.
The converse implication is similar:
if $u(s,t)>0$ then by definition of matrix product,
there exists $q\in Q$ such that 
$u_1(s,q)>0$ and $u(q,t)>0$, and we use the induction hypothesis to
get $\limu_+(s,t)=1$. This concludes the proof of~\eqref{eq:induc1}.
Now we prove that~\eqref{eq:induc2} holds.
Let $s,t\in Q$ such that $\limu(s,t)=1$.
By definition of a decomposition tree,
$\limu=\limu_{1}\cdot \limu_{2}$.
Since $\limu(s,t)=1$ then by definition
of the product of two limit-words there exists $q\in Q$ such that
$\limu_{1}(s,q)=1$ and $\limu_{2}(q,t)=1$.
Then $u(s,t)\geq u_1(s,q) \cdot u_2(q,t)\geq\pmin^{2^{h}} \cdot \pmin^{2^{h}}=
\pmin^{2^{h+1}}$ where the first inequality is by definition of the matrix
product and the second inequality is by induction hypothesis.
This completes the proof of~\eqref{eq:induc2}.

{\bf Second case, $t$ is an iteration node} labelled
by $(u,(\limu^\sharp,\limu_+))$ with $k$ sons $t_1,\ldots,t_k$ labelled
by $(u_1,(\limu,\limu_{+})),\ldots,(u_k,(\limu,\limu_{+}))$.
The proof that~\eqref{eq:induc1} holds is similar to the concatenation
node case (and relies on the fact that $\limu_+$ is idempotent).
We focus on the proof of~\eqref{eq:induc2}.
Let $s,r\in Q$ such that $\limu^\sharp(s,r)=1$.
By definition of a decomposition tree,
$\limu=\limu_{1}\cdots \limu_{k}$.
Since $t$ is an iteration node,
$k\geq 3$ thus:
\begin{equation}\label{eq:grou}
u(s,r) \geq u_1(s,r) \cdot \sum_{q\in Q} \left( \left(u_2\cdots u_{k-1}\right)(r,q) \cdot u_k(q,r) \right)\enspace.
\end{equation}

To establish~\eqref{eq:induc2} we prove that:
\begin{align}
\label{eq:grou1}&
u_1(s,r)\geq \pmin^{2^{h}},
\\
\label{eq:grou2}&
\forall q\in Q, (u_2\cdots u_{k-1})(r,q)>0 \implies u_k(q,r)\geq\pmin^{2^{h}}.
\end{align}

First we prove~\eqref{eq:grou1}.
Since $\limu^\sharp(s,r)=1$ then by definition of the iteration operation,
$r$ is $\limu$-recurrent and $\limu(s,r)=1$.
By induction hypothesis applied to $t_1$, according to~\eqref{eq:induc2},
it implies $u_1(s,r)\geq\pmin^{2^{h}}$ {\it i.e}~\eqref{eq:grou1}.

Now we prove~\eqref{eq:grou2}.
For that we use the hypothesis that $(\limu,\limu_+)$ is not a leak witness.
Let $q\in Q$ such that $(u_2\cdots u_{k-1})(r,q)>0$.
Then by induction hypothesis applied to $t_2,\ldots,t_{k-1}$,
according
to~\eqref{eq:induc1}, $\limu_{+}^{k-2}(r,q)=1$.
Thus by idempotency of
$\limu_+$, $\limu_{+}(r,q)=1$. Since
by hypothesis $\limu^\sharp(s,r)=1$
then $r$ is $\limu$-recurrent
and since $(\limu,\limu_+)$ is not a leak witness
then necessarily $\limu(q,r)=1$.
Thus, by induction hypothesis and according to~\eqref{eq:induc2},
$u_k(q,r)\geq\pmin^{2^{h}} $ {\it i.e}~\eqref{eq:grou2}.

Now, putting~\eqref{eq:grou},~\eqref{eq:grou1} and~\eqref{eq:grou2} altogether,
\begin{align*}
u(s,r)&\geq u_1(s,r) \cdot \sum_{q\in Q} (u_2\cdots u_{k-1})(r,q) \cdot u_k(q,r)\\
&\geq \pmin^{2^{h}} \cdot \sum_{q\in Q} (u_2\cdots u_{k-1})(r,q) \cdot \pmin^{2^{h}}\\
&\geq \pmin^{2^{h+1}} ,
\end{align*}
where the second inequality holds because $\sum_{q\in Q} (u_2\cdots
u_{k-1})(r,q)=1$ by basic property of transition matrices.
This completes the proof of~\eqref{eq:induc2}.

\medskip

To conclude, according to~\eqref{eq:depth}
the depth of a decomposition tree can be bounded by $3 \cdot
|\monoid_+|^2$,
and since $\monoid_+$ has less than $\numjclasses = 2^{2|Q|^2}$ elements
the depth $h$ is less than $3 \cdot
\numjclasses^2$.
Then according to~\eqref{eq:induc1} and~\eqref{eq:induc2}
this completes the proof of Lemma~\ref{lem:lowerbound}.
\end{IEEEproof}

\section{\label{sec:ex}A few leaktight automata}

In this section, we present several properties and examples of leaktight automata.

\subsection{Two basic examples}
\begin{figure}[ht]
\begin{center}
\includegraphics[scale=.8]{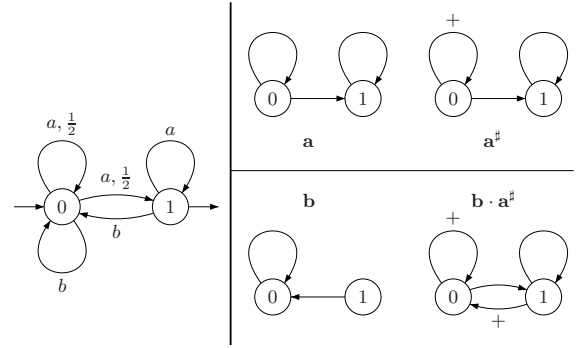}
\caption{\label{fig:3} A leaktight automaton and its extended Markov monoid.}
\end{center}
\end{figure}

The automaton on Fig.~\ref{fig:3} is leaktight.
Its extended Markov monoid is depicted on the right-hand side (except for the neutral element $(\mathbf 1,\mathbf 1)$).
Each of the four directed graphs represents an extended limit-word $(\limu,\limu_+)$;
the edges marked $+$ are the edges that are in $\limu_+$ but not in $\limu$.

The initial state of the automaton is state $0$, and the unique final state is state $1$.
This automaton has value $1$ and this can be checked using its Markov monoid: 
there is a single value $1$ witness $\lima^\sharp$.
Notice that the two distinct extended limit-words labelled by $\lima^\sharp$ and
$\limb \cdot \lima^\sharp$ on Fig.~\ref{fig:3} correspond to the same limit-word $\lima^\sharp$.

\begin{figure}[ht]
\begin{center}
\includegraphics[scale=.8]{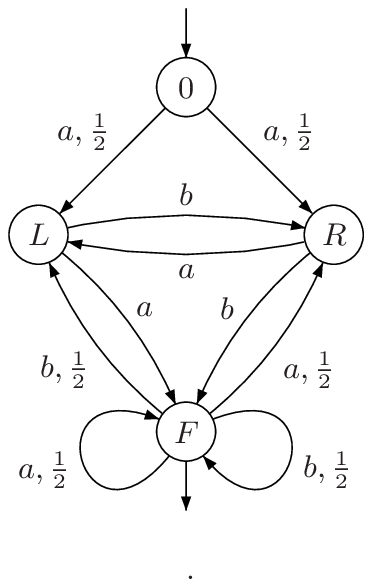}
\caption{\label{fig:4} A leaktight automaton which does not have value $1$.}
\end{center}
\end{figure}

The automaton on Fig.~\ref{fig:4} is leaktight.
The initial state of the automaton is state $0$, and the unique final state is state $F$.
The Markov monoid has too many elements to be represented here.
This automaton does not have value $1$.

\vspace{-0.5cm}

\subsection{The class of leaktight automata is rich and stable}

The class of leaktight automata contains all known classes of probabilistic automata with a decidable
value $1$ problem, in particular hierarchical automata defined
in~\cite{ChadhaSV09} and $\sharp$-acyclic automata defined in~\cite{GO10}.

\begin{proposition}\label{prop:containment}
Deterministic automata, hierarchical probabilistic automata and
$\sharp$-acyclic automata are leaktight and these inclusions are strict.
\end{proposition}

Another interest of the class of leaktight automata
is its stability under two natural composition operators:
parallel composition and synchronized product.
An automaton $\AA||\BB$
is the parallel composition
of two automata $\AA$ and $\BB$
if its state space is the disjoint union
of the state spaces of $\AA$ and $\BB$
plus a new initial state.
For every input letter, the possible successors of the
initial state are itself or one of the
initial state of $\AA$ and $\BB$.
An automaton $\AA\times\BB$
is the synchronized product
of two automata $\AA$ and $\BB$
if its state space is the cartesian product
of the state spaces of $\AA$ and $\BB$,
with induced transition probabilities.

\begin{proposition}\label{prop:stable}
The leaktight property is stable by parallel composition
and synchronized product.
\end{proposition}

\subsection{About $\sharp$-height}

\begin{figure}[ht]
\begin{center}
\includegraphics[scale=.7]{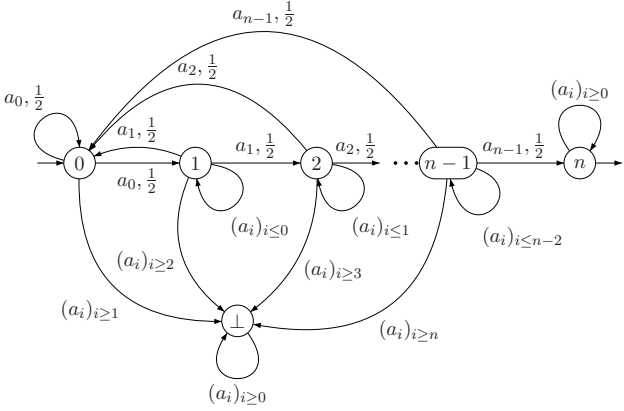}
\caption{\label{fig:sharpstar} A leaktight automaton with value $1$ and
$\sharp$-height $n$.}
\end{center}
\end{figure}

The $\sharp$-height of an automaton 
is the maximum over all elements $\limu$ of its Markov monoid
of the minimal number of nested applications of the iteration operator needed to
obtain $\limu$.
As already mentioned, an adaptation of a result by Kirsten (Lemma 5.7 in~\cite{Kirsten05})
shows that the $\sharp$-height of an automaton is at most $|Q|$.
A natural question is whether this bound is tight. The answer is positive,
a simple computation shows that the only value $1$ witness of the
automaton of Fig.~\ref{fig:sharpstar} is $\limu=(\cdots((\lima_0^\sharp
\lima_1)^\sharp \lima_2)^\sharp \lima_3)^\sharp \cdots \lima_{n-1})^\sharp$,
whose $\sharp$-height is $n = |Q|-2$.

\vspace{-0.5cm}

%A result similar to Proposition~\ref{prop:flat}
%holds for hierarchical automata.
The following proposition shows a crucial difference
between leaktight automata
and $\sharp$-acyclic automata.
%otherthe other known classes of
%automata with a decidable value $1$ problem.

\begin{proposition}\label{prop:flat}
Deterministic automata and
$\sharp$-acyclic automata have $\sharp$-height $1$.
\end{proposition}

\section*{Conclusion}
We introduced a subclass of probabilistic automata, called leaktight automata,
for which we proved that the value $1$ problem is $\PSPACE$-complete.

In the present paper we considered automata over finite words.
Next step is the adaptation of our results to infinite words 
and probabilistic B{\"u}chi automata~\cite{BBG08,ChadhaSV09},
as well as partially observable Markov decision processes.

A natural question is ``what does the Markov monoid say about a probabilistic automaton (leaktight or not)?''.
Since the Markov monoid is independent of
the actual values of transition probabilities (only positivity matters),
this suggests the two following questions.
Given a probabilistic automaton whose transition probabilities are unspecified
(only positivity is specified),
\begin{enumerate}
  \item is it decidable whether the answer to the value $1$ problem is
the same for \emph{any} choice of transition probabilities?
  \item does the Markov monoid contain a value $1$ witness if and only if
the automaton has value $1$ for \emph{some} choice of transition probabilities?
\end{enumerate}
The first question, suggested by a referee of the present paper, is open,
while the answer to the second question seems to be negative.

\section*{Acknowledgment}
We thank Thomas Colcombet for having pointed us to the work of Leung and Simon,
as well as two referees for their careful reading and their
constructive comments and help in improving this paper.
\bibliographystyle{plain}
\bibliography{bib}

\onecolumn

\appendix

\newenvironment{appendixtheorem}[2][Theorem]{\begin{trivlist}
\item[\hskip \labelsep {\bfseries #1}\hskip \labelsep {\bfseries #2}]\begin{em}}{\end{em}\end{trivlist}}

\newenvironment{appendixlemma}[2][Lemma]{\begin{trivlist}
\item[\hskip \labelsep {\bfseries #1}\hskip \labelsep {\bfseries
#2}]\begin{em}}{\end{em}\end{trivlist}}

\newenvironment{appendixproposition}[2][Proposition]{\begin{trivlist}
\item[\hskip \labelsep {\bfseries #1}\hskip \labelsep {\bfseries
#2}]\begin{em}}{\end{em}\end{trivlist}}

\section*{Tool lemmata}

We start with a few tool lemmata.

\begin{appendixlemma}{\ref{lem:idempotenteverywhere}}
The following two properties hold:
\begin{itemize}
	\item For every word $u\in A^*$,
the word $u^{|Q|!}$ is idempotent.
	\item For every limit-word $\limu\in\{0,1\}^{Q^2}$,
the limit-word $\limu^{|Q|!}$ is idempotent.
\end{itemize}
\end{appendixlemma}

\begin{IEEEproof}
The second statement implies the first one, so we prove the second one.

Let $n = |Q|$ and $s,t\in Q$ such that $\limu^{n!}(s,t)=1$.
We want to prove that $\limu^{2 \cdot n!}(s,t)=1$.
Since $\limu^{n!}(s,t)=1$,
there exists $q \in Q$ and $k,l<|Q|$
such that $\limu^k(s,q)=1$,
$\limu^{n!-k-l}(q,q)=1$
and
$\limu^l(q,t)=1$.
Consequently, there exists
$k'<|Q|$ such that $\limu^{k'}(q,q)=1$,
and since $k' | n!$, this implies
$\limu^{n!}(q,q) = 1$, thus $\limu^{2 \cdot n! - k - l}(q,q)=1$
and finally $\limu^{2 \cdot n!}(s,t)=1$.

The proof that $\limu^{2 \cdot n!}(s,t) = 1$ implies $\limu^{n!}(s,t)=1$ is similar.
\end{IEEEproof}

The following lemma provides two simple yet useful properties.

\begin{lemma}\label{lem:trivial}
Let $\AA$ be a probabilistic automaton.
\begin{enumerate}
  \item [i)]
Let $\limu$ be an idempotent limit-word.
Then for each state $s\in Q$, there exists $t\in Q$
such that $\limu(s,t)=1$ and $t$ is $\limu$-recurrent.
\item[ii)]
Let $(\limv,\limv_+)$ be an extended limit-word of the extended Markov monoid of $\AA$.
Then:
\begin{equation}\label{eq:trivial}
\forall s,t\in Q, (\limv(s,t)=1 \implies \limv_+(s,t)=1)\enspace.
\end{equation}
\end{enumerate}
\end{lemma}

\begin{IEEEproof}
We prove i).
Let $C\subseteq Q$ be a strongly connected component of the graph $\limu$ reachable from $s$. 
(Recall that $\limu$ can be seen as the directed graph whose vertex set is $Q$ and where there is an edge from $s$ to $t$
if $\limu(s,t) = 1$.)
Then for every $t,q\in C$ there exists $k$
and a path $t=t_1,t_2,\ldots, t_k=q$ from $t$ to $q$ in $\limu$.
Thus, $(\limu^k)(t,q)=1$ and since $\limu$ is idempotent,
$\limu(t,q)=1$. Thus, $C$ is a clique of the graph $\limu$ and all states of $C$ are recurrent.
Since $C$ is reachable from $s$ in $\limu$, the same argument proves that for every $t\in C$, $\limu(s,t)=1$.

We prove ii).
By definition, the extended Markov monoid is the smallest monoid containing
$\set{(\lima,\lima) \mid a \in A}$ and $(\mathbf 1,\mathbf 1)$, stable by concatenation and iteration.
Property~\eqref{eq:trivial} holds for every pair $(\lima,\lima)$
where $a\in A$. Moreover this property is stable by concatenation and
iteration. This completes the proof.
\end{IEEEproof}

\section*{The value $1$ problem depends quantitatively on the transition probabilities}

In this section, we give a short proof of the following lemma, claimed about Fig~\ref{fig:x}:
\begin{lemma}
The probabilistic automaton $\AA$ has value $1$ if $x > \frac{1}{2}$.
\end{lemma}

We first note:
$$\prob{\AA}(0 \xrightarrow{b a^n} \top) = \frac{1}{2} \cdot x^n \qquad 
\textrm{ and } \qquad \prob{\AA}(0 \xrightarrow{b a^n} \bot) = \frac{1}{2} \cdot (1-x)^n$$

Fix an integer $N$ and consider the following stochastic process: 
it consists in (at most) $N$ identical rounds.
In a round, there are three outcomes: winning with probability $p_n = \frac{1}{2} \cdot x^n$,
losing with probability $q_n = \frac{1}{2} \cdot (1-x)^n$, or a draw with probability $1 - (p_n + q_n)$.
Once the game is won or lost, it stops, otherwise it goes on to the next step, until the $N$\textsuperscript{th} round.
This stochastic process mimics the probabilistic automaton reading the input word $(b a^n)^N$.

The probability to win is:
$$\begin{array}{lll}
\prob{}(\textrm{Win}_N) & = & \sum_{k = 1}^N \prob{}(\textrm{Win at round } k) \\[1.5ex]
& = & \sum_{k = 1}^N (1 - (p_n + q_n))^{k-1} \cdot p_n \\[1.5ex]
& = & p_n \cdot \frac{1 - (1 - (p_n + q_n))^N}{1 - (1 - (p_n + q_n))} \\[1.5ex]
& = & \frac{p_n}{p_n + q_n} \cdot (1 - (1 - (p_n + q_n))^N) \\[1.5ex]
& = & \frac{1}{1 + \frac{q_n}{p_n}} \cdot (1 - (1 - (p_n + q_n))^N) \\[1.5ex]
\end{array}$$

We now set $N = n$.
A simple calculation shows that the sequence $(1 - (1 - (p_n + q_n))^n)_{n \in \NN}$ converges to $1$ as $n$ goes to infinity.
Furthermore, if $x > \frac{1}{2}$ then $\frac{1-x}{x} < 1$, 
so $\frac{q_n}{p_n} = \left(\frac{1-x}{x}\right)^n$ converges to $0$ as $n$ goes to infinity.
It follows that the probability to win converges to $1$ as $n$ goes to infinity.
Consequently: $$\lim_n \prob{\AA}(0 \xrightarrow{(b a^n)^n} \top) = 1\ .$$

\section*{Undecidability in a very restricted case}

In this section we sharpen the undecidability result of the value $1$ problem to probabilistic automata
having only one probabilistic transition.

We say that a probabilistic automaton is \textit{simple} if for all letters $a$, for all states $s$ and $t$,
we have $M_a(s,t) \in \set{0,\frac{1}{2},1}$.
%Hugo: I dont agree with the following statement
%Any probabilistic automaton can
% be easily transformed into a simple automaton, so the value $1$ problem is undecidable for simple probabilistic automata.
According to~\cite{GO10},
the value $1$ problem is undecidable for simple probabilistic automata.
We first show how to simulate a (simple) probabilistic automaton with one having only one probabilistic transition, 
\textit{up to a regular language}:
\begin{proposition}
Given a simple probabilistic automaton $\AA = (Q,A,(M_a)_{a \in A}, q_0, F)$, 
there exists a simple probabilistic automaton $\BB$ over a new alphabet $B$, with one probabilistic transition,
and a morphism $\widehat{\_} : A^* \longrightarrow B^*$ such that:
$$\forall w \in A^*, \prob{\AA}(w) = \prob{\BB}(\widehat{w}).$$
\end{proposition}

The morphism $\widehat{\_}$ will not be onto, so this simulation works 
up to the regular language $\widehat{A^*} = \set{\widehat{w} \mid w \in A^*}$.
We shall see that the automaton $\BB$ will not be able to check that a word read belongs 
to this language, which makes this restriction unavoidable in this construction.

We first give the intuitions behind the construction.
Intuitively, while reading the word $w$, the probabilistic automaton $\AA$ 
``throw parallel threads''.
A computation of $\AA$ over $w$ can be viewed as a tree, 
where probabilistic transitions correspond to branching nodes.

\begin{figure}[ht]
\begin{center}
\includegraphics[scale=.8]{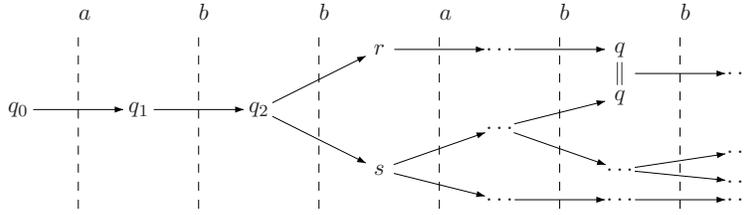}
\caption{\label{ex_computation} An example of a computation.}
\end{center}
\end{figure}

On the figure, reading $a$ from $q_0$ or $b$ from $q_1$ leads deterministically to the next state.
Reading $b$ from $q_2$ leads at random to $r$ or to $s$, hence the corresponding node is branching. 
Our interpretation is that two parallel threads are thrown.
Let us make two observations:
\begin{itemize}
	\item threads are not synchronized: reading the fourth letter (an $a$), 
	the first thread leads deterministically to the next state, while the second thread randomizes;
	\item threads are merged so there are at most $n = |Q|$ parallel threads: 
	whenever two threads synchronize to the same state $q$, they are merged.
	This happens in the figure after reading the fifth letter ($b$).
\end{itemize}

The automaton $\BB$ we construct will simulate the $n$ threads from the beginning, 
and take care of the merging process each step.

\begin{IEEEproof}
We denote by $q_i$ the states of $\AA$, \textit{i.e} $Q = \set{q_0, \ldots, q_{n-1}}$.
The alphabet $B$ is made of two new letters `$*$' and `$\merge$' plus, 
for each letter $a \in A$ and state $q \in Q$,
two new letters $\chck(a,q)$ and $\apply(a,q)$, so that:
$$B = \set{*, \merge} \cup \bigcup_{a \in A, q \in Q} \set{\chck(a,q), \apply(a,q)}$$

We now define the automaton $\BB$.
We duplicate each state $q \in Q$, and denote the fresh copy by $\bar{q}$.
Intuitively, $\bar{q}$ is a temporary state that will be merged at the next merging process.
States in $\BB$ are either a state from $Q$ or its copy, or one of the three fresh states $s_*$, $s_0$ and $s_1$.

The initial state remains $q_0$ as well as the set of final states remains $F$.

The transitions of $\BB$ are as follows:
\begin{itemize}
	\item for every letter $a \in A$ and state $q \in Q$, 
the new letter $\chck(a,q)$ from state $q$ leads deterministically to state $s_*$ \textit{i.e} $M_{\chck(a,q)}(q) = s_*$,
	\item the new letter $*$ from state $s_*$ leads with probability half to $s_0$ and half to $s_1$, 
	\textit{i.e} $M_{s_*}(*) = \frac{1}{2} s_0 + \frac{1}{2} s_1$ (this is the only probabilistic transition of $\BB$);
	\item the new letter $\apply(a,q)$ from states $s_0$ and $s_1$ applies the transition function from $q$ reading $a$: 
if the transition $M_a(q)$ is deterministic, \textit{i.e} $M_a(q,r) = 1$ for some state $r$ 
then $M_{\apply(a,q)}(s_0) = \bar{r}$ and $M_{\apply(a,q)}(s_1) = \bar{r}$,
else the transition $M_a(q)$ is probabilistic \textit{i.e}
$M_a(q) = \frac{1}{2} r + \frac{1}{2} r'$ for some states $r,r'$, 
then $M_{\apply(a,q)}(s_0) = \bar{r}$ and $M_{\apply(a,q)}(s_1) = \bar{r'}$;
	\item the new letter $\merge$ activates the merging process: it consists in replacing $\bar{q}$ by $q$ for all $q \in Q$.
\end{itemize}
Whenever a couple (letter, state) does not fall in the previous cases, it has no effect.
The gadget simulating a transition is illustrated in the figure.

\begin{figure}[ht]
\begin{center}
\includegraphics[scale=.8]{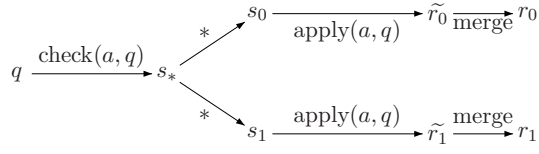}
\caption{\label{gadget1} The first gadget.}
\end{center}
\end{figure}

Now we define the morphism $\widehat{\_} : A^* \longrightarrow B^*$ by its action on letters:
$$\widehat{a} =
\chck(a,q_0) \cdot * \cdot \apply(a,q_0) \ldots \chck(a,q_{n-1}) \cdot * \cdot \apply(a,q_{n-1}) \cdot \merge.$$

The computation of $\AA$ while reading $w$ in $A^*$
is simulated by $\BB$ on $\widehat{w}$, \textit{i.e} we have:
$$\prob{\AA}(w) = \prob{\BB}(\widehat{w})$$

This completes the proof.
\end{IEEEproof}

Let us remark that $\BB$ is indeed unable to check that a letter $\chck(a,q)$ 
is actually followed by the corresponding $\apply(a,q)$: 
in-between, it will go through $s_*$ and ``forget'' the state it was in.

We now improve the above construction: we get rid of the regular external condition.
To this end, we will use probabilistic automata whose transitions have probabilities $0$, $\frac{1}{3}$, $\frac{2}{3}$ or $1$.
This is no restriction, as stated in the following lemma:

\begin{lemma}
For any simple probabilistic automaton $\AA = (Q,A,(M_a)_{a \in A}, q_0, F)$, 
there exists a probabilistic automaton $\BB$ whose transitions have probabilities $0$, $\frac{1}{3}$, $\frac{2}{3}$ or $1$, 
such that for all $w$ in $A^*$, we have:
$$\val{\AA} = \val{\BB}.$$
\end{lemma}

\begin{IEEEproof}
We provide a construction to pick with probability half, using transitions with probability
$0$, $\frac{1}{3}$, $\frac{2}{3}$ and $1$.
The construction is illustrated in the figure.

\begin{figure}[ht]
\begin{center}
\includegraphics[scale=.8]{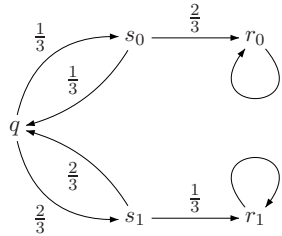}
\caption{\label{gadget} Simulating a half with a third.}
\end{center}
\end{figure}

In this gadget, the only letter read is a fresh new letter $\sharp$.
The idea is the following: to pick with probability half $r_0$ or $r_1$, 
we sequentially pick with probability a third or two thirds.
Whenever the two picks are different, if the first was a third, 
then choose $r_0$, else choose $r_1$. 
This happens with probability half each.
We easily see that $\prob{\AA}(a_0 \cdot a_1 \cdot \ldots a_{k-1}) 
= \sup_p \prob{\BB}(a_0 \cdot \sharp^p \cdot a_1 \cdot \sharp^p \ldots a_{k-1} \cdot 
\sharp^p)$.

\end{IEEEproof}

\begin{proposition}
For any simple probabilistic automaton $\AA = (Q,A,(M_a)_{a \in A}, q_0, F)$, 
there exists a simple probabilistic automaton $\BB$ over a new alphabet $B$, 
with one probabilistic transition, such that:
$$\val{\AA} = 1 \Longleftrightarrow \val{\BB} = 1\ .$$
\end{proposition}

Thanks to the lemma, we assume that in $\AA$, transitions have probabilities $0$, $\frac{1}{3}$, $\frac{2}{3}$ or $1$.
The new gadget used to simulate a transition is illustrated in the figure.

\begin{figure}[ht]
\begin{center}
\includegraphics[scale=.8]{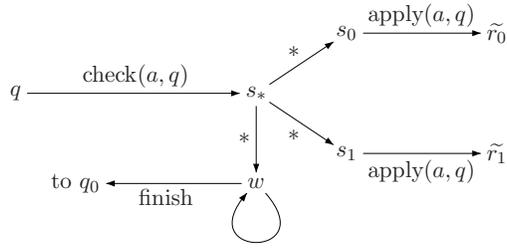}
\caption{\label{gadget2} The second gadget.}
\end{center}
\end{figure}

The automaton $\BB$ reads words of the form $u_1 \cdot \finish \cdot u_2 \cdot \finish \ldots$, 
where `$\finish$' is a fresh new letter.
The idea is to ``skip'', or ``delay'' part of the computation of $\AA$:
each time the automaton $\BB$ reads a word $u_i$, it will be skipped with some probability.

Simulating a transition works as follows: 
whenever in state $s_*$, reading the letter `$*$' leads 
with probability one third to $s_1$, one third to $s_0$ and one third to $w$.
As before, from $s_0$ and $s_1$, we proceed with the simulation.
However, in the last case, we ``wait'' for the next letter `$\finish$' 
that will restart from $q_0$.
Thus each time a transition is simulated, the word being read is skipped 
with probability $\frac{1}{3}$.

Delaying part of the computation allows to multiply the number of threads.
We will use the accepted threads to check the extra regular condition we had before.
To this end, as soon as a simulated thread is accepted in $\BB$, 
it will go through an automaton (denoted $\CC$ in the construction) 
that checks the extra regular condition.

\begin{IEEEproof}
We keep the same notations.
The alphabet $B$ is made of three new letters: `$*$', `$\merge$' and `$\finish$' plus, 
for each letter $a \in A$ and state $q \in Q$,
two new letters $\chck(a,q)$ and $\apply(a,q)$, so that:
$$B = \set{*, \merge, \finish} \cup \bigcup_{a \in A, q \in Q} \set{\chck(a,q), \apply(a,q)}$$

We first define a syntactic automaton $\CC$.
We define a morphism $\widehat{\_} : A^* \longrightarrow B^*$ by its action on letters:
$$\widehat{a} =
\chck(a,q_0) \cdot * \cdot \apply(a,q_0) \ldots \chck(a,q_{n-1}) 
\cdot * \cdot \apply(a,q_{n-1}) \cdot \merge.$$
Consider the regular language $L = \set{\widehat{w} \cdot \finish \mid w \in A^*}^*$, 
and $\CC = (Q_\CC,\delta_\CC,s_\CC,F_\CC)$ an automaton recognizing it.

We now define the automaton $\BB$.
We duplicate each state $q \in Q$, and denote the fresh copy by $\bar{q}$.
States in $\BB$ are either a state from $Q$ or its copy, 
a state from $Q_\CC$ or one of the four fresh states $s_*$, $s_0$, $s_1$ and $\wait$.

The initial state remains $q_0$, and the set of final states is $F_\CC$.

The transitions of $\BB$ are as follows:
\begin{itemize}
	\item for every letter $a \in A$ and state $q \in Q$, 
the new letter $\chck(a,q)$ from state $q$ leads deterministically to state $s_*$ \textit{i.e} $M_{\chck(a,q)}(q) = s_*$,
	\item the new letter $*$ from state $s_*$ leads with probability 
one third to $s_*$, one third to $s_0$ and one third to $s_1$, 
\textit{i.e} $M_{s_*}(*) = \frac{1}{3} s_* + \frac{1}{3} s_0 + \frac{1}{3} s_1$ 
(this is the only probabilistic transition of $\BB$);
	\item any other letter from state $s_*$ leads deterministically to $w$,
\textit{i.e} $M_{s_*}(\_) = \wait$;
	\item the new letter $\apply(a,q)$ from states $s_0$ and $s_1$ applies 
the transition function from $q$ reading $a$: 
if the transition $M_a(q)$ is deterministic, 
\textit{i.e} $M_a(q,r) = 1$ for some state $r$ then $M_{\apply(a,q)}(s_0) = \bar{r}$ 
and $M_{\apply(a,q)}(s_1) = \bar{r}$,
else the transition $M_a(q)$ is probabilistic \textit{i.e}
$M_a(q) = \frac{1}{2} r + \frac{1}{2} r'$ for some states $r,r'$, 
then $M_{\apply(a,q)}(s_0) = \bar{r}$ and $M_{\apply(a,q)}(s_1) = \bar{r'}$;
	\item the new letter $\merge$ activates the merging process: 
it consists in replacing $\bar{q}$ by $q$ for all $q \in Q$;
	\item the new letter $\finish$ from state $\wait$ leads deterministically to $q_0$;
	\item the new letter $\finish$ from state $q$ in $F$ leads deterministically to $s_\CC$;
	\item the new letter $\finish$ from any other state is not defined (there is a deterministic transition to a bottom non-accepting state).
\end{itemize}
Transitions in $\CC$ are not modified.
Whenever a couple (letter, state) does not fall in the previous cases, it has no effect.

We now show that this construction is correct.

We first prove that for all $w \in A^*$, there exists a sequence of words $(w_p)_{p \geq 1}$ 
such that $\prob{\AA}(w) = \sup_p \prob{\BB}(w_p)$.

The probability to faithfully simulate one transition is $\left(\frac{2}{3}\right)^n$.
It follows:
$$\delta_\BB (q_0,\hat{w} \cdot \finish) = 
\left(\frac{2}{3}\right)^k \prob{\AA}(w)
+ \left(1 - \left(\frac{2}{3}\right)^k\right) q_0,$$
for some $k$ satisfying $n \leq k \leq n \cdot |w|$ (the number $k$ corresponds to the number of transitions followed
along a faithful simulation of $w$).
%The computation of $\AA$ while reading $w$ is simulated by $\BB$ on $\widehat{w} \cdot \finish$. 
This implies that $\sup_p \prob{\BB}((\widehat{w} \cdot \finish)^p) = \prob{\AA}(w)$,
hence if $\val{\AA} = 1$, then $\val{\BB} = 1$.

Conversely, we prove that if $\val{\BB} = 1$, then $\val{\AA} = 1$.
Let $w$ a word read by $\BB$ accepted with probability close to $1$, we factorize it as follows: 
$w = u_1 \cdot \finish \cdot \ldots \cdot u_k \cdot \finish$, 
such that $u_i$ does not contain the letter $\finish$.
The key observation is that if $k = 1$, 
the word $w$ is accepted with probability at most $\frac{2}{3}$. 
Hence we consider only the case $k > 1$.
We assume without loss of generality that $\prob{\AA}(u_1) > 0$ 
(otherwise we delete $u_1 \cdot \finish$ and proceed).
In this case, a thread has been thrown while reading $u_1$ that reached $s_\CC$, 
so the syntactic process started: 
it follows that $u_i$ for $i > 1$ are in the image of $\widehat{\_}$.
This implies that the simulation is sound: 
from $w$ we can recover a word in $A^*$ accepted 
with probability arbitrarily close to $1$ by $\AA$.

This completes the proof.
\end{IEEEproof}

The proposition implies the following corollary:
the value $1$ problem is undecidable, even for probabilistic automata with only one probabilistic transition.

\section*{Decomposition trees of bounded span}

\begin{appendixtheorem}{\ref{theo:decomposition}}
Let $A$ be a finite alphabet, $(M,\cdot,1)$ a monoid equipped with a
function $\sharp$ that maps every idempotent $\lime\in M$
to another idempotent element $\lime^\sharp\in M$
and $\phi: A^* \to M$ a morphism.
Every word $u \in A^+$ has a decomposition tree whose span is less than 
$3 \cdot |M|$.
\end{appendixtheorem}

The following proof appeared in~\cite{simontropical}; we give it here for the sake of completeness.
\begin{IEEEproof}
We start by adding a few letters to $A$.
For every idempotent $\lime \in M$,
we add a letter $\underline{\lime}$ to the alphabet $A$, and extend $\phi$ by 
$\phi(\underline{\lime}) = \lime$.
We do not lose generality because this operation
does not modify the monoid $M$,
and a decomposition tree for a word with letters from the original alphabet
cannot use the new letters of the extended alphabet $\underline{A} = A \cup
\{\underline{\lime} \mid \lime \in M, \lime=\lime^2\}$.

We proceed by induction on the length of $u$.

First, if $u$ is a letter $a$, the decomposition tree whose 
only node is labelled by $(a,\phi(a))$ has span $1$.

Consider now a word $u$ with at least two letters.
According to
Theorem~\ref{theo:simon},
there exists a Ramseyan factorization tree $T_u$ of $u$ of height less than $3
\cdot |M|$.
According to Lemma~\ref{lem:decomp},
this factorization tree is in general not a decomposition tree,
except in the very special case
where it has no discontinuous nodes.
The rest of the proof shows how to transform $T_u$ into
a decomposition tree of
span less than $3\cdot|M|$.
The proof technique consists in taking care of the discontinuous nodes
of $T_u$
in a bottom-up fashion.

If $T_u$ has no discontinuous node then it is already a decomposition tree.
Moreover its span is equal to its height, which is less than $3\cdot|M|$.

If $T_u$ has at least one discontinuous node,
let $t$ be such a node with maximal depth.
By definition, $t$ has at least three children $t_1,\ldots,t_k$
labelled by $(v_1,\phi(v_1)),\ldots,(v_k,\phi(v_k))$ and $t$ itself is labelled
by $(v,\phi(v)) = (v_1 \cdots v_k ,\phi(v_1 \cdots v_k))$.
We have $\phi(v_1) =
\ldots = \phi(v_k) = e$ and since $t$ is discontinuous,
$e \neq e^\sharp$.
We distinguish two cases.
\begin{itemize}
	\item Either $t$ is the root of $T_u$. In that case $v = u$
and we can construct directly a decomposition tree $T_u$ of $u$ 
of span less than $3 \cdot |M|$.
Let $T_1,\ldots,T_k$ be the subtrees of $T_u$ whose roots are respectively 
$t_1,\ldots,t_k$.
Then, since $t$ is a discontinuous node of maximal depth in $T_u$, 
each subtree $T_1,\ldots,T_k$ contains no discontinuous node at
all. Consequently, each subtree $T_i$ is a decomposition tree
whose span is equal to its height, which is less than $3 \cdot |M|-1$.
Then a decomposition tree for $u$ is the tree 
with the root labelled by $(u,e^\sharp)$
and children $T_1,\ldots,T_k$.
Since $e \neq e^\sharp$, the root is discontinuous and 
the span of this tree is less than $3 \cdot |M|$.
	\item Or $t$ is not the root of $T_u$.
Since $t$ is labelled by $(v,e^\sharp)$, 
there exist two words $w,w' \in A^+$ such that $u = w \cdot v \cdot w'$.
We replace the subword $v$ in $u$ by the letter $\underline{\lime^\sharp}$
and obtain the word $u' = w \cdot \underline{\lime^\sharp} \cdot w'$.
Since $t$ is a discontinuous node, it is an iteration node and
has at least three
children, thus $v$ has length at least $3$. 
Thus, $u'$ is strictly shorter than $u$ and we can apply 
the induction hypothesis to $u'$: 
let $T'$ be a decomposition tree for $u'$, whose span is less than $3 \cdot
|M|$. One of the leaves of $T'$ corresponds to the letter
$\underline{\lime^\sharp}$ of $u'$ and is labelled by
$(\underline{\lime^\sharp},\lime^\sharp)$. 
We replace this leaf by the decomposition tree $T_v$ of $v$ given by induction hypothesis. 
Since the root of $T_v$ is labelled by $(v,\lime^\sharp)$,
we obtain a decomposition tree for $u = w \cdot v \cdot w'$, 
whose span is less than $3 \cdot |M|$.
This completes the induction step.
\end{itemize}
\end{IEEEproof}

\begin{appendixlemma}{\ref{lem:discont}}
Let $A$ be a finite alphabet, and $M$ a monoid equipped with a
function $\sharp$ that maps every idempotent $e\in M$
to another idempotent element $e^\sharp\in M$.
Suppose moreover that for every idempotent $e \in M$,
\begin{equation}
e^\sharp \cdot e =
e^\sharp = e \cdot e^\sharp\enspace.
\end{equation}

Then for every idempotent element $e\in M$,
either $e^\sharp=e$ or $e^\sharp<_\JJ e$.

As a consequence, the number of discontinuous nodes along
a path in a decomposition tree is at most $\numjclasses$,
where $\numjclasses$ is the number of $\mathcal{J}$-classes of the monoid.
\end{appendixlemma}

\begin{IEEEproof}
We prove the first part of the lemma.
Equation~\eqref{eq:esharp}
implies that $e^\sharp = e^\sharp e e^\sharp$ thus $e^\sharp \leq_\JJ e$.
Now, we suppose that $e \leq_\JJ e^\sharp$ and prove that $e=e^\sharp$.
Since $M$ is finite, we have $e \DD e^\sharp$.
Since $e \cdot e^\sharp = e^\sharp$, it follows $e \RRR e^\sharp$.
By a dual argument, we have $e \LL e^\sharp$; hence $e \HH e^\sharp$.
Both $e$ and $e^\sharp$ and idempotents, so according to Green's theorem
(see \textit{e.g}~\cite{howie}, Theorem 2.2.5.)
$e = e^\sharp$.

The second part of the lemma is an immediate consequence, 
since the sequence of elements of the monoid labelling a branch of a decomposition
tree, starting from the root,
is non-decreasing for the $\JJ$-order and strictly
increasing on discontinuous nodes.
Indeed, if an internal node has two children labelled by 
$(u_1,\limu_1)$ and $(u_2,\limu_2)$ then by definition of a decomposition tree
the node is labelled by $(u_1 \cdot u_2,\limu_1 \cdot \limu_2)$
and by definition of the $\JJ$-order,
$\limu_1 \cdot \limu_2 \leq_\JJ \limu_1$ and $\limu_1 \cdot \limu_2 \leq_\JJ \limu_2$.
For continuous nodes, the same idempotent labels both the internal node and its
children. For discontinuous nodes, the father node is labelled by some idempotent
$\lime^\sharp$ and the children by $\lime \neq \lime^\sharp$, and we conclude
with supra.
\end{IEEEproof}

\section*{About the extended Markov monoid and leak witnesses}

\begin{appendixtheorem}{\ref{theo:leakwitness}}
An automaton $\AA$ is leaktight if and only if its
extended Markov monoid contains no leak witness.
\end{appendixtheorem}

The proof is split in two parts, the direct implication (Lemma~\ref{lem:leakwitnessdirect})
and the converse implication (Lemma~\ref{lem:leakwitnessconverse}).

\begin{lemma}\label{lem:leakwitnessdirect}
If the extended Markov monoid of an automaton $\AA$
contains a leak witness then $\AA$ has a leak.
\end{lemma}

\begin{IEEEproof}
Suppose that there is a leak witness $(\limu,\limu_+)$
in the extended Markov monoid.

By definition of a leak witness,
$\limu$ and $\limu_+$ are idempotent and there exists $r,q\in Q$
such that $r$ is $\limu$-recurrent, $\limu_+(r,q)=1$ and $\limu(q,r)=0$. 
We prove now that there exists a leak from $r$ to $q$.

By induction, following the proof of Lemma~\ref{lem:consistency}
for each $(\limu,\limu_+)$ in the extended Markov monoid,
we build a sequence $(u_n)_{n\in\NN}$ such that
for every states $s,t\in Q$
the sequence $(u_n(s,t))_{n\in\NN}$ converges and
\begin{align}
\label{eq:appcond1}
&\limu(s,t)=1\iff \lim_n u_n(s,t)>0\enspace,\\
\label{eq:appcond2}
&\limu_+(s,t)=0\iff \forall n\in\NN, u_n(s,t)=0\enspace,\\
\label{eq:appcond3}
&\limu_+(s,t)=1\iff \forall n\in\NN, u_n(s,t)>0\enspace.
\end{align} 

To complete the proof,
we show that $(u_n)_{n\in\NN}$ is a leak in $\AA$ from $r$ to $q$.
According to Definition~\ref{def:leak}, there are four conditions to be met.

First condition is by hypothesis.
Moreover, since $\limu_+$ is idempotent then according to~\eqref{eq:appcond1}
all the $(u_n)_{n\in\NN}$ are idempotent.
Second, let $\mathcal{M}_u$
the Markov chain associated with transition probabilities $(u(s,t))_{s,t\in Q}$.
We prove that $r$ is $\mathcal{M}_u$-recurrent.
Since $r$ is $\limu$-recurrent, and according to~\eqref{eq:appcond1},
$\forall s\in Q, u(r,s)>0 \implies u(s,r)>0$.
But $u$ is idempotent because $\limu$ is idempotent and~\eqref{eq:appcond1}.
Thus according to Lemma~\ref{lem:idempotent}, $r$ is $u$-recurrent.
Third, we need to prove $\forall n\in\NN, u_n(r,q)>0$, this holds
because
of~\eqref{eq:appcond3}
and $\limu_+(r,q)=1$.
Fourth, we need to prove that $r$ is not reachable from $q$ in
$\mathcal{M}_u$.
Since $u=\lim_n u_n$ and according to~\eqref{eq:appcond1},
$\limu(s,t)=1\iff u(s,t)>0$,
thus accessibility in the directed graph $\limu$ and
in the Markov chain $\MM_u$ coincide.
Since $\limu(r,q)=0$ and $\limu$ is idempotent,
$r$ is not accessible from $q$ in $\limu$,
thus neither accessible in $\mathcal{M}_u$.
\end{IEEEproof}

\begin{lemma}\label{lem:leakwitnessconverse}
If the extended Markov monoid of an automaton $\AA$
contains no leak witness then $\AA$ is leaktight.
\end{lemma}

\begin{IEEEproof}
By contraposition,
suppose there is a leak $(u_n)\nNN$ from a state $r$
to a state $q$ in $\AA$,
and for each $s,t\in q$,
denote $u(s,t)$ the limit of 
the sequence $(u_n(s,t))\nNN$
and $\mathcal{M}_u$ the Markov chain induced by $(u(s,t))_{s,t\in Q}$.
By definition of a leak:
\begin{align}
\label{eq:leak1}
%  \item[1)] 
&\text{$r$ is recurrent in $\mathcal{M}_u$},\\
\label{eq:leak2}
%  \item[2)] 
&\forall n\in \NN, u_n(r,q)>0,\\
\label{eq:leak3}
%  \item[3)] 
&r \text{ is not accessible from $q$ in $\mathcal{M}_u$}.
\end{align}

To get to the conclusion,
we use the leak $(u_n)\nNN$
to build a leak witness $(\limv,\limv_+)$ in the extended
monoid $\monoid_+$ of $\AA$.

The first task is to define the pair $(\limv,\limv_+)$.
By hypothesis, we can apply the Lemma~\ref{lem:lowerbound}
to each word $u_n$ of the leak,
which gives for each $n\in\NN$ a pair $(\limu_n,\limu_{+,n})\in\monoid_+$
such that for all states $s,t$:
\begin{align}
\label{eq:tt1}
\limu_{+,n}(s,t)=1 &\iff u_n(s,t)>0\enspace,\\
\label{eq:tt2}
\limu_n(s,t) = 1 &\implies  u_n(s,t) \geq \dabound \enspace.
\end{align}
Since the extended Markov monoid is finite,
there exists $N\in\NN$
such that:
\begin{equation}
\label{eq:infoften}
\text{for infinitely many }
n\in\NN, (\limu_N,\limu_{+,N})=(\limu_n,\limu_{+,n})\enspace.
\end{equation}
Let $(\limv,\limv_+) = (\limu_N,\limu_{+,N})^{|\monoid_+|!}$.
Then according to Lemma~\ref{lem:idempotenteverywhere}, $(\limv,\limv_+)$
is idempotent.
Note also that according to~\eqref{eq:tt1}
and since the $u_n$ are idempotent (by definition of leaks),
 \begin{equation}\label{eq:limvplus}
 \limu_{+,N}\text{ is idempotent and } \limv_+ = \limu_{+,N}\enspace.
 \end{equation}

Now, we prove that $(\limv,\limv_+)$ is a leak witness.
According to i) of Lemma~\ref{lem:trivial},
since $\limv$ is idempotent, there exists
$r'$ such that $\limv(r,r')=1$ and $r'$ is $\limv$-recurrent.
By definition of a leak witness, if we prove that
(a) $\limv_+(r',q)=1$, (b) $\limv(q,r')=0$ then
$(\limv,\limv_+)$ is a leak witness.

We first prove (a).
Let $\eta=\dabound$ and $K=|\monoid_+|!$.
Then:
\begin{align}
\notag&\limv(r,r')=1
&\text{(by definition of $r'$)}\\
\notag&\implies \limu_N^K(r,r')=1
&\text{(by definition of $\limv$)}\\
\notag&\implies \limu_n^K(r,r')=1, \text{ for infinitely many $n$}
&\text{(by definition of $N$)}\\
\notag&\implies u_n^K(r,r')\geq \eta^K, \text{ for infinitely many $n$}
&\text{(by~\eqref{eq:tt2})}\\
\notag&\implies u^K(r,r')\geq \eta^K
&\text{(because $u=\lim_n u_n$)}\\
\notag& \implies \text{$r'$ is $u$-recurrent}
&\text{(because $r$ is $u$-recurrent)}\\
\label{eq:rprimer}&\implies \exists l, u^l(r',r)>0\\
\notag&\text{(because $r$ and $r'$ are in the same class of $u$-recurrence)}\\
\notag&\implies \exists l, u^{l+1}(r',q)>0
&\text{(because $u(r,q)>0)$}\\
\notag&\implies \exists l, \exists N',\forall n\geq N', u_n^{l+1}(r',q)>0
&\text{(because $u=\lim_n u_n$)}\\
\notag&\implies \exists N',\forall n\geq N', u_n(r',q)>0
&\text{(the $u_n$ are idempotent)}\\
\notag&\implies \exists N',\forall n\geq N',\limu_{+,n}(r',q)=1
&\text{(by~\eqref{eq:tt1})}\\
\notag&\implies \limu_{+,N}(r',q)=1
&\text{(by definition of $N$)}\\
\notag&\implies \limv_{+}(r',q)=1
&\text{(according to~\eqref{eq:limvplus})}.
\end{align}
  
Now we prove (b).
By contradiction, suppose that $\limv(q,r')=1$.
Then:
\begin{align}
\notag&\limv(q,r')=1\\
\notag&\implies \limu_N^K(q,r')=1
&\text{(by definition of $\limv$)}\\
\notag&\implies \limu_n^K(q,r')=1, \text{ for infinitely many $n$}
&\text{(by definition of $N$)}\\
\notag&\implies u_n^K(q,r')\geq \eta^K, \text{ for infinitely many $n$}
&\text{(by~\eqref{eq:tt2})}\\
\notag&\implies u^K(q,r')\geq \eta^{K},
&\text{(because $u=\lim_n u_n$)}\\
\notag&\implies \text{$r'$ is reachable from $q$ in $\mathcal{M}_u$}
&\text{}\\
\notag&\implies \text{$r$ is reachable from $q$ in $\mathcal{M}_u$}
&\text{(according to~\eqref{eq:rprimer})}
% 
% &\text{(by~\eqref{eq:tt2})}\\
% &\implies u^K(r,r')\geq \eta^K
% &\text{(by continuity)}\\
% 
% &\implies \limv^2(r,q)=1
% &\text{( because $\limv(r,r')=1$)}\\
% &\implies \limv(r,q)=1
% &\text{( because $\limv$ idempotent )}\\
% &\implies \limu_N^K(r,q)=1
% &\text{( by definition of $\limv$ )}\\
% &\implies \limu_n^K(r,q)=1 \text{ for infinitely many $n\in\NN$}
% &\text{( by definition of $N$ )}\\
% &\implies u_n^K(r,q)>\eta^K \text{ for infinitely many $n\in\NN$}
% &\text{( by~\eqref{eq:tt2} )}\\
% %&\implies u_n(r,q)>\eta^K \text{ for infinitely many $n\in\NN$}
% %&\text{( the $u_n$ are idempotent)}\\
% &\implies u^K(r,q)>\eta^K
% &\text{( by continuity )},
\end{align}
which contradicts~\eqref{eq:leak3}.

This completes the proof of (b),
thus $(\limv,\limv_+)$ is a leak witness,
which concludes the proof of Lemma~\ref{lem:leakwitnessconverse}.
%According to Lemma~\ref{lem:weakcompleteness},
%the Markov monoid $\monoid$ of $\AA$ is complete,
%hence there exists a limit-word $\limu\in \monoid$
%\begin{equation}
%\lim u_n(s,t)=0 \implies u(s,t)=0\enspace.
%\end{equation}
\end{IEEEproof}

\section*{About $\sharp$-height}

In this section, we adapt Kirsten's results to show that the $\sharp$-height of an automaton is at most $|Q|$.
The following statements are straightforward translations
from~\cite{Kirsten05}, we provide them for the sake of completeness.

Consider $\limu$ an idempotent limit-word. We define $\sim_\limu$ the relation on $Q$ by
$i \sim j$ if $\limu(i,j) = 1$ and $\limu(j,i) = 1$.
Clearly, $\sim_\limu$ is symmetric, and since $\limu$ is idempotent, $\sim_\limu$ is transitive.
If for some $i$ there is a $j$ such that $i \sim_\limu j$, then $i \sim_\limu i$ thanks to $\limu$'s idempotency.
Consequently, the restriction of $\sim_\limu$ to the set
$$Z_\limu = \set{i \in Q \mid \textrm{ there is some } j \textrm{ such that } i \sim_\limu j}$$
is reflexive, \textit{i.e} $\sim_\limu$ is an equivalence relation on $Z_\limu$.
By equivalence class of $\sim_\limu$ we mean an equivalence class of $\sim_\limu$ on $Z_\limu$.
We denote by $[i]_\limu$ the equivalence class of $i$, 
and by $\Cl(\limu)$ the set of equivalence classes of $\sim$.

\begin{lemma}\label{lem_tool}
The following two properties hold:
\begin{itemize}
	\item Let $\limu,\limv$ be two limit-words and $i,j$ in $Q$.
Then $(\limu \cdot \limv)(i,j) \geq \limu(i,k) \cdot \limv(k,j)$ for all $k$ in
$Q$.
	\item Let $\limu$ be an idempotent limit-word and $i,j$ in $Q$.
There is some $l$ in $Q$ such that $\limu(i,j) = \limu(i,l) \cdot \limu(l,l) \cdot \limu(l,j)$.
\end{itemize}
\end{lemma}

\begin{IEEEproof}
The first claim is clear and follows from the equality:
$$(\limu \cdot \limv)(i,j) = \sum_{k \in Q} \limu(i,k) \cdot \limv(k,j)\ .$$
Consider now the second claim. For every $k$, we have
$\limu(i,j) = (\limu^3)(i,j) = \sum_{l,l' \in Q} \limu(i,l) \cdot \limu(l,l') \cdot \limu(l',j) \geq
\limu(i,k) \cdot \limu(k,k) \cdot \limu(k,j)$.
Since $\limu = \limu^{n+2}$, there are $i = i_0,\ldots,i_{n+2} = j$ such that 
$\limu(i,j) = \limu(i_0,i_1) \cdot \ldots \cdot \limu(i_{n+1},i_{n+2})$. 
By a counting argument, there are $1 \leq p < q \leq (n + 1)$ such that $i_p = i_q$. 
Let $l = i_p$. 
We have 
$$\begin{array}{c}
\limu(i,l) = \limu^p(i,l) \geq \limu(i_0,i_1) \cdot \ldots \cdot \limu(i_{p-1},i_p), \\
\limu(l,l) = \limu^{q-p}(l,l) \geq \limu(i_p,i_{p+1}) \ldots \limu(i_{q-1},i_q), \textrm{ and } \\
\limu(l,j) = \limu^{n+2-q}(l,j) \geq \limu(i_q,i_{n+2}) \ldots \limu(i_{n+1},i_{n+2})\ .
\end{array}$$
Hence, $\limu(i,l) \cdot \limu(l,l) \cdot \limu(l,j) \geq \limu(i_0,i_1) \ldots \limu(i_{p-1},i_p) = \limu(i,j)$,
and the second claim follows.
\end{IEEEproof}

\begin{appendixlemma}{\ref{lem:rec_class_inclu_strict}}
Let $\limu$ and $\limv$ be two idempotent limit-words. Assume $\limu \leq_\JJ \limv$,
then $|\Cl(\limu)| \leq |\Cl(\limv)|$.
\end{appendixlemma}

\begin{IEEEproof}
%e = v et f = u
Let $\lima$, $\limb$ two limit-words such that $\lima \cdot \limv \cdot \limb = \limu$. 
We assume $\lima \cdot \limv = \lima$ and $\limv \cdot \limb = \limb$. 
If $\lima$ and $\limb$ do not satisfy these conditions, 
then we proceed the proof for $\lima = \lima \cdot \limv$ and $\limb = \limv \cdot \limb$.
We construct a partial surjective mapping $\beta : \Cl(\limv) \rightarrow \Cl(\limu)$. 
The mapping $\beta$ depends on the choice of $\lima$ and $\limb$. 
For every $i,j$ with $i \sim_\limv i$ and $j \sim_\limu j$ 
satisfying $\lima(j, i) \cdot \limv(i, i) \cdot \limb(i, j) = 1$,
we set $\beta([i]_\limv) = [j]_\limv$. 
To complete the proof, we have to show that $\beta$ is well defined and that $\beta$ is indeed surjective.

We show that $\beta$ is well defined. Let $i,i'$ such that $i \sim_\limv i$ and $i' \sim_\limv i'$. 
Moreover, let $j,j'$ such that $j \sim_\limu j$ and $j' \sim_\limu j'$. 
Assume $\lima(j,i) \cdot \limv(i,i) \cdot \limb(i,j) = 1$ and 
$\lima(j',i') \cdot \limv(i',i') \cdot \limb(i',j') = 1$.
Thus, $\beta([i]_\limv) = [j]_\limu$ and $\beta([i']_\limv) = [j']_\limu$. 
To show that $\beta$ is well defined, we have to show that if
$[i]_\limv = [i']_\limv$, then $[j]_\limu = [j']_\limu$. 
Assume $[i]_\limv = [i']_\limv$, \textit{i.e}, $i \sim_\limv i'$.
Hence, $\limv(i,i') = 1$.
Above, we assumed $\lima(j,i) \cdot \limv(i,i) \cdot \limb(i,j) = 1$, and thus, 
$\lima(j,i) = 1$. 
Similarly, $\limb(i',j') = 1$.
Consequently, $\lima(j,i) \cdot \limv(i,i') \cdot \limb(i',j') = 1$, \textit{i.e}, 
$\limu(j,j') = (\lima \cdot \limv \cdot \limb)(j,j') = 1$. 
By symmetry, we achieve $\limu(j',j) = 1$, and hence, $j \sim_\limu j'$.

We show that $\beta$ is surjective. Let $j$ such that $j \sim_\limu j$. We have to exhibit some $i$ such that
$\beta([i]_\limv) = [j]_\limu$. 
Since $j \sim_\limu j$, we have $\limu(j,j) = 1$. 
Since $\limu = \lima \cdot \limv \cdot \limb$, there are $k, l$ such that
$\lima(j,k) \cdot \limv(k,l) \cdot \limb(l,j) = 1$, 
and in particular, $\limv(k,l) = 1$. 
By Lemma~\ref{lem_tool} there is some $i$ such
that $\limv(k,i) \cdot \limv(i,i) \cdot \limv(i,l) = \limv(k,l) = 1$, 
and in particular, $\limv(i,i) = 1$. 
We have $\lima(j,i) = (\lima \cdot \limv)(j,i) \geq \lima(j,k) \cdot \limv(k,i) = 1$, 
and $\limb(i,j) = (\limv \cdot \limb)(i,j) \geq \limv(i,l) \cdot \limb(l,j) = 1$. 
To sum up, $\lima(j,i) \cdot \limv(i,i) \cdot \limb(i,j) = 1$,
and hence, $\beta([i]_\limv) = [j]_\limu$.
\end{IEEEproof}

\begin{appendixlemma}{\ref{lem:rec_class_inclu}}\label{lem_st}
Let $\limu$ be an idempotent limit-word. We have $\Cl(\limu^\sharp) \subseteq \Cl(\limu)$.
Furthermore, if $\limu^\sharp \neq \limu$ then $\Cl(\limu^\sharp) \subsetneq \Cl(\limu)$.
\end{appendixlemma}

\begin{IEEEproof}
Let $i$ be such that $i \sim_{\limu^\sharp} i$. We show $[i]_\limu = [i]_{\limu^\sharp}$.
For every $j$ with $i \sim_{\limu^\sharp} j$, we have, $i \sim_\limu j$. Hence, $[i]_{\limu^\sharp} \subseteq [i]_\limu$.
Conversely, let $j \in [i]_\limu$; we have $\limu(i,j) = 1$. 
Since $i \sim_{\limu^\sharp} i$, we have $\limu^\sharp(i,i) = 1$. 
To sum up, $\limu^\sharp(i,j) = (\limu^\sharp \cdot \limu)(i,j) \geq \limu^\sharp(i,i) \cdot \limu(i,j) = 1$,
and by symmetry, $\limu^\sharp(j,i) = 1$, \textit{i.e}, $i \sim_{\limu^\sharp} j$.
Hence $j \in [i]_{\limu^\sharp}$.

Assume now $\limu^\sharp \neq \limu$. 
Let $i,j$ such that $\limu(i,j) = 1$ and $\limu^\sharp(i,j) = 0$. By Lemma~\ref{lem_tool}, 
there is some $l$ such that $\limu(i,j) = \limu(i,l) \cdot \limu(l,l) \cdot \limu(l,j)$,
so $\limu(l,l) = 1$.
By contradiction, assume $\limu^\sharp(l,l) = 1$. Hence,
$$\limu^\sharp(i,j) = (\limu \cdot \limu^\sharp \cdot \limu)(i,j) \geq 
\limu(i,l) \cdot \limu^\sharp(l,l) \cdot \limu(l,j)
= \limu(i,l) \cdot \limu(l,l) \cdot \limu(l,j) = 1,$$
\textit{i.e}, $\limu^\sharp(i,j) = 1$ which is a contradiction. 
Consequently, $\limu^\sharp(l,l) = 0$, so $l \sim_\limu l$ and $l \not\sim_{\limu^\sharp} l$. 
Thus, $l \in Z_\limu$ but $l \notin Z_{\limu^\sharp}$. 
Hence, there is a class $[l]_\limu$ in $\Cl(\limu)$, 
but there is no class $[l]_{\limu^\sharp}$ in $\Cl(\limu^\sharp)$. 
In combination with the first part of this lemma, we obtain $\Cl(\limu^\sharp) \subsetneq \Cl(\limu)$.
\end{IEEEproof}

\section*{A few leaktight automata}

\begin{appendixproposition}{\ref{prop:containment}}
Deterministic automata, hierarchical probabilistic automata and
$\sharp$-acyclic automata are leaktight.
\end{appendixproposition}
\begin{IEEEproof}
It is obvious that deterministic automata are leaktight.
We give an algebraic proof.
For deterministic automata the iteration operation
has no effect on limit-words.
As a consequence, the extended Markov monoid only contains
pair $(\limu,\limu)$ whose both components are equal,
and none of them can be a leak witness.
The characterization given by Theorem~\ref{theo:good},
allows us to conclude that deterministic automata are leaktight.

The proof for hierarchical automata is given
in Proposition~\ref{prop:hierarchical}.

The proof for $\sharp$-acyclic automata is given
in Proposition~\ref{prop:sharpacyclic}.
\end{IEEEproof}

\begin{appendixproposition}{\ref{prop:stable}}
The leaktight property is stable by parallel composition
and synchronized product.
\end{appendixproposition}
\begin{IEEEproof}
Both cases are proved easily.

For the parallel product,
let $i$ be the new initial state.
If there is a leak in $\AA||\BB$
from a state $q\neq i$
then this a leak either in $\AA$
or $\BB$.
There can be no leak $(u_n)\nNN$ from $i$
because $i$ is $u$-recurrent only for those words
$u$ that are written with letters stabilizing $i$.

For the synchronized product,
the extended Markov monoid of the synchronized product $\AA\times\BB$
is the product of the extended Markov monoids of $\AA$
and $\BB$.
If there was a leak in
$\AA\times\BB$, then according to Theorem~\ref{theo:leakwitness}
there would be a leak witness
$(\limu,\limu_+)= ((\limu_\AA,\limu_\BB),(\limu_{+,\AA},\limu_{+,\BB}))$ in
the extended Markov monoid of $\AA\times\BB$
from a state $(r_\AA,r_\BB)$ to a state $(q_\AA,q_\BB)$.
Then $r_\AA$ is $\limu_\AA$-recurrent and $\limu_{+,\AA}(r_\AA,q_\AA)=1$
thus since $\AA$ is leaktight $\limu_{\AA}(q_\AA,r_\AA)=1$.
Similarly, $\limu_{\BB}(q_\BB,r_\BB)=1$
thus $\limu((q_\AA,q_\BB),(r_\AA,r_\BB))=1$
hence a contradiction. 
\end{IEEEproof}

\subsection{Leaktight automata strictly contain hierarchical automata}

The class of hierarchical automata has been defined
in~\cite{ChadhaSV09}.

The states $Q$ of a hierarchical automaton are sorted according to levels
such that for each letter, at most one successor is at the same level and all
others
are at higher levels.
\newcommand{\rank}{\text{rank}}
Formally, there is a mapping $\rank:Q\to[1,\ldots,l]$
such that
$\forall a\in A, \forall s,t\in Q$ such that $a(s,t)>0$,
$\rank(t)\geq \rank(s)$ and the set $\{t\mid a(s,t)>0, \rank(t)=\rank(s)\}$
is either empty or a singleton.

\begin{proposition}\label{prop:hierarchical}
Every hierarchical automata is leaktight.
\end{proposition}

%\begin{IEEEproof}
%We show by contradiction that there is no leak witness in the extended Markov
%monoid of a hierarchical automata.
%Let $(\limu,\limu_+)$ be such a witness, from state $r$ to state $q$.
%Since $\limu(q,r)=1$, according to ii) of Lemma~\ref{lem:trivial},
%$\limu_+(q,r)=1$. Thus, there exists a finite word $u\in A^*$
%such that $u(q,r)>0$. Therefore, since $q\neq r$,
%$\rank(q)<\rank(r)$ thus for every letter $a\in A$,
%$a(q,r)=0$ thus $\limu_+(q,r)=0$, a contradiction
%with the definition of a leak.
%\end{IEEEproof}

\begin{IEEEproof}
We prove by induction that for every extended limit-word
$(\limu,\limu_+)$
in the extended Markov
monoid of a hierarchical automata,
for every state $r$:
\begin{equation}
\label{eq:hier}
(\text{$r$ is $\limu$-recurrent}) \implies (\forall q\neq r, \limu_+(r,q)=
0)\enspace.
\end{equation}

Property~\eqref{eq:hier} obviously holds for base elements $(\lima,\lima)$.

Property~\eqref{eq:hier} is stable by product:
let $(\limu,\limu_+)$ and $(\limv,\limv_+)$
with property~\eqref{eq:hier} and let $r\in Q$
be $\limu \limv$-recurrent.
By definition of hierarchical automata
the recurrence classes of the limit-words $\limu,\limv$ and $\limu\limv$
are singletons
%then $\forall q\in Q, (\limu\limv)(r,q)=1\implies q=r$
thus $r$ is necessarily both $\limu$-recurrent and $\limv$-recurrent.
According to~\eqref{eq:hier},
$\forall q\neq r$, $\limu_+
(r,t)=0$ thus $\forall q\neq r, (\limu_+\limv_+)(r,q)= 0$.

Property~\eqref{eq:hier} is obviously stable by iteration,
which terminates the proof.
\end{IEEEproof}

The inclusion is strict, an example is given by Fig.~\ref{fig:3}.

\subsection{Leaktight automata strictly contain $\sharp$-acyclic automata}

The class of $\sharp$-acyclic automata has been defined in~\cite{GO10}.

Let $\AA$ be a probabilistic automaton, to define $\sharp$-acyclic automata, 
we define an action on non-empty subsets of states.
Given $S \subseteq 2^Q$ and a letter $a$, by definition
$S \cdot \lima = \set{t \mid \exists s \in S, \lima(s,t) = 1}$.
If $S \cdot \lima = S$, then we define the iteration of $\lima$:
$S \cdot \lima^\sharp = \set{t \mid \exists s \in S, \lima^\sharp(s,t) = 1}$.
Consider now the graph whose vertices are non-empty subsets of states 
and there is an edge from $S$ to $T$ if
$S \cdot \lima = T$ or $S \cdot \lima = S$ and $S \cdot \lima^\sharp = T$.
The automaton $\AA$ is $\sharp$-acyclic if the unique cycles in this graph are self loops.

We extend the action on any limit-word: given $S \subseteq Q$ and 
a limit-word $\limu$, by definition
$S \cdot \limu = \set{t \mid \exists s \in S, \limu(s,t) = 1}$.

% \begin{definition}
% $\AA$ is not $\sharp$-acyclic if and only if there exists $S,T$ subsets of states,
% $S \neq T$, $\limu,\limv$ two limit-words such that 
% $S \cdot \limu = T$ and $T \cdot \limv = S$.
% \end{definition}

\begin{appendixproposition}{\ref{prop:flat}}
Deterministic automata,
$\sharp$-acyclic automata and hierarchical automata have $\sharp$-height $1$.
\end{appendixproposition}
\begin{IEEEproof}
For deterministic automata, this is obvious
because there are no unstable idempotent in the Markov monoid
so the iteration operation is useless and the $\sharp$-height
is actually $0$.

For $\sharp$-acyclic automata,
this is a corollary 
of results in~\cite{GO10}:
if a $\sharp$-acyclic automaton has value $1$
then there exists a sequence of letters $a_0,b_0,a_1,\ldots,a_n,b_n,a_{n+1}\in
(A\cup\{\epsilon\})^*$ such that
$a_0b_0^\sharp a_1b_1^\sharp\ldots
a_nb_n^\sharp a_{n+1}$
is a value $1$ witness.
\end{IEEEproof}

\begin{proposition}\label{prop:sharpacyclic}
Every $\sharp$-acyclic automata is leaktight.
\end{proposition}

\begin{IEEEproof}
We prove that for all extended limit-word $(\limu,\limu_+)$, we have
$(\limu(s,t) = 0, \limu_+(s,t)=1) \Rightarrow s \textrm{ is transient}$,
which implies the leaktight assumption, by induction on $\limu$.
The case $\limu = \lima$ is clear.
Consider the case $\limu = \limv^\sharp$, and let $s,t$ states 
such that $(\limu(s,t) = 0, \limu_+(s,t)=1)$.
Then either $(\limv(s,t) = 0, \limv_+(s,t)=1)$ or $\limv(s,t) = 1$ 
and $t$ is transient in $\limv$.
In the first case, the induction hypothesis ensures that $s$ is transient in $\limv$.
In the second case, $s$ would be transient in $\limv$.
In both cases, $s$ is transient in $\limv$, so also in $\limu$.

Consider now the case $\limu = \limu_1 \cdot \limu_2$,
and let $s,t$ states such that $(\limu(s,t) = 0, \limu_+(s,t)=1)$.
Assume toward contradiction that $s$ is recurrent in $\limu$.
Let $C = \set{q \mid \limu(s,q) = 1}$ be the recurrence class of $s$,
so we have $C \cdot \limu = C$.
The $\sharp$-acyclicity implies that $C \cdot \limu_1 = C$ and $C \cdot \limu_2 = C$.

There are two cases: either there exists $p$ such that $(\limu_1(s,p) = 0, \limu_{1_+}(s,p)=1)$,
or such that $\limu_1(s,p) = 1$ and $(\limu_{2}(s,p) = 0, \limu_{2_+}(s,p)=1)$.
Consider the first case, and $T = C \cdot \limu_1^\sharp$.
We have $T \subsetneq C$, so $T \cdot u^\sharp = C$, which defines a $\sharp$-cycle over $C$,
contradiction.
In the second case, let $T = C \cdot \limu_2^\sharp$, we have $T \subsetneq C$ so
$T \cdot \limu^\sharp = C$, which defines a $\sharp$-cycle over $C$, contradiction.
This completes the proof.
\end{IEEEproof}

The inclusion is strict:
Fig.~\ref{fig:sharpstar} provides an example of
leaktight automaton which is not $\sharp$-acyclic.

\end{document}